\documentclass[conference]{IEEEtran}
\IEEEoverridecommandlockouts
\usepackage[letterpaper,%
            left=0.625in,right=0.625in,top=0.75in,bottom=1in]{geometry}
\usepackage{cite}
\usepackage{amsmath,amssymb,amsfonts}
\usepackage{algorithm}
\usepackage[noend,compatible]{algpseudocode}
\usepackage{graphicx}
\usepackage{textcomp}
\usepackage{xcolor}
\usepackage{fancyhdr}
\usepackage[hyphens]{url}
\usepackage{subfig}
\usepackage{soul}
\usepackage{caption}
\usepackage{comment}
\usepackage{booktabs}
\usepackage{tabularx}
\usepackage{kotex}
\usepackage{multirow}
\usepackage{tikz}
\usepackage{balance}

\usepackage{xfrac}
\usepackage{makecell}
\usepackage{array}
\usepackage{siunitx}
\DeclareRobustCommand{\encircled}[1]{\raisebox{.5pt}{\textcircled{\raisebox{-.9pt}{#1}}}}
\usepackage{hyperref}
\hypersetup{
     nolinks=true,
}

\makeatletter
    \newcommand*{\algrule}[1][\algorithmicindent]{\makebox[#1][l]{\hspace*{.5em}\thealgruleextra\vrule height \thealgruleheight depth \thealgruledepth}}%
\newcommand*{\thealgruleextra}{}
\newcommand*{\thealgruleheight}{.90\baselineskip}
\newcommand*{\thealgruledepth}{.25\baselineskip}

\newcount\ALG@printindent@tempcnta
\def\ALG@printindent{%
    \ifnum \theALG@nested>0
        \ifx\ALG@text\ALG@x@notext
        \else
            \unskip
            \addvspace{-1pt}
            \ALG@printindent@tempcnta=1
            \loop
                \algrule[\csname ALG@ind@\the\ALG@printindent@tempcnta\endcsname]%
                \advance \ALG@printindent@tempcnta 1
            \ifnum \ALG@printindent@tempcnta<\numexpr\theALG@nested+1\relax
            \repeat
        \fi
    \fi
    }%
\usepackage{etoolbox}
\patchcmd{\ALG@doentity}{\noindent\hskip\ALG@tlm}{\ALG@printindent}{}{\errmessage{failed to patch}}
\makeatother

\newbox\statebox
\newcommand{\myState}[1]{%
    \setbox\statebox=\vbox{#1}%
    \edef\thealgruleheight{\dimexpr \the\ht\statebox+1pt\relax}%
    \edef\thealgruledepth{\dimexpr \the\dp\statebox+1pt\relax}%
    \ifdim\thealgruleheight<.75\baselineskip
        \def\thealgruleheight{\dimexpr .75\baselineskip+1pt\relax}%
    \fi
    \ifdim\thealgruledepth<.25\baselineskip
        \def\thealgruledepth{\dimexpr .25\baselineskip+1pt\relax}%
    \fi
    \State #1%
    \def\thealgruleheight{\dimexpr .75\baselineskip+1pt\relax}%
    \def\thealgruledepth{\dimexpr .25\baselineskip+1pt\relax}%
}

\def\BibTeX{{\rm B\kern-.05em{\sc i\kern-.025em b}\kern-.08em
    T\kern-.1667em\lower.7ex\hbox{E}\kern-.125emX}}

\title{Accelerating Number Theoretic Transformations for Bootstrappable Homomorphic Encryption on GPUs}
\author{
  \IEEEauthorblockN{Sangpyo Kim, Wonkyung Jung, Jaiyoung Park, and Jung Ho Ahn}
  \IEEEauthorblockA{Seoul National University, Seoul, Republic of Korea\\
  \{vnb987, jungwk, jeff1273, gajh\}@snu.ac.kr}
  \thanks{\copyright~2020 IEEE. Personal use of this material is permitted.  Permission from IEEE must be obtained for all other uses, in any current or future media, including reprinting/republishing this material for advertising or promotional purposes, creating new collective works, for resale or redistribution to servers or lists, or reuse of any copyrighted component of this work in other works.}
}

\begin{document}
\maketitle
\pagestyle{empty}


\begin{abstract}
  Homomorphic encryption (HE) draws huge attention as it provides a way of 
  privacy-preserving computations on encrypted messages.
  Number Theoretic Transform (NTT), a specialized form of Discrete Fourier 
  Transform (DFT) in the finite field of integers, is the key algorithm that
  enables fast computation on encrypted ciphertexts in HE.
  Prior works have accelerated NTT and its inverse transformation on a popular
  parallel processing platform, GPU, by leveraging DFT optimization techniques.
  However, these GPU-based studies lack a comprehensive analysis of the primary
  differences between NTT and DFT or only consider small HE parameters that 
  have tight constraints in the number of arithmetic operations that can be
  performed without decryption.

  In this paper, we analyze the algorithmic characteristics of NTT and DFT
  and assess the performance of NTT when we apply the optimizations that are
  commonly applicable to both DFT and NTT on modern GPUs.
  From the analysis, we identify that NTT suffers from severe main-memory 
  bandwidth bottleneck on large HE parameter sets.
  To tackle the main-memory bandwidth issue, we propose a novel NTT-specific 
  on-the-fly root generation scheme dubbed on-the-fly twiddling (OT).
  Compared to the baseline radix-2 NTT implementation, after applying all the 
  optimizations, including OT, we achieve 4.2$\times$ speedup on a modern GPU.
\end{abstract}

\section{Introduction}
\label{sec:introduction}
In the current cloud computing era, users exploit a convenient environment in which to access cloud resources easily.
However, this convenience poses a new privacy concern.
To access cloud computing environments, a user should send private data to the cloud server.
However, such processes increase the possibility of leaking private data.

Homomorphic encryption (HE)~\cite{fsc-1978-he} has been highlighted as a safe way to solve this privacy problem.
HE is a cryptographic scheme that enables computation on encrypted messages (called \emph{ciphertexts}).
The results of computing on ciphertexts followed by decryption are identical to those when applying the computation to corresponding unencrypted messages.
With HE, a user encrypts her data in a private space while saving the key for decryption.
The cloud server can perform a series of computations on the encrypted data.
Even if the cloud server becomes compromised, the intruder cannot decrypt the encrypted data without the private key; hence, the privacy of user's data is still ensured.

Ciphertext in HE is represented as multiple polynomials whose coefficients are represented as big integers.
The magnitude of the big integer is bounded by the ciphertext modulus Q.
However, performing arithmetic operations on big integers is inefficient.
Typical HE schemes~\cite{cheon-2018-full,han-2020-better,bajard-2016-full} use the Chinese remainder theorem (CRT) to avoid arithmetic operations with big integers.
When considering a polynomial as a sequence of coefficients, CRT converts a sequence of coefficients into multiple sequences of residues, where each sequence has a different modulus.
The length of each sequence is identical to the degree of the polynomial, 
and the number of sequences is identical to the number of primes needed to represent a big integer uniquely.

The number Theoretic Transform (NTT)~\cite{mc-1965-ntt} is a specialized form of the Discrete Fourier Transform (DFT) in the finite field of integers.
DFT uses the powers of the $N_{th}$ root of unity (i.e., $e^{(-2\pi j/N)}$) as twiddle factors 
when converting a sequence of complex numbers with length $N$. 
In contrast, NTT uses the powers of the $N_{th}$ root of unity modulo a prime number as twiddle factors to perform modular arithmetic operations in an integer space.

NTT is a crucial enabling algorithm for fast HE computations~\cite{jung-2020-heaan,roy-2019-fpga}.
For example, in one study~\cite{roy-2019-fpga}, NTT and its inverse (iNTT) account for 34\% of the entire ciphertext multiplication process
under the condition of the degree of the polynomials in ciphertext being $2^{12}$ and the ciphertext modulus $Q=2^{180}$.
In addition, NTT/iNTT consumes 50.04\% of the total processing time when undertaking the multiplication of ciphertext
under the condition of the degree of the polynomials in the ciphertext being $2^{15}$ and the ciphertext modulus $Q=2^{881}$ using the open-source library SEAL~\cite{icfcds-2017-seal} on the CPU.
A number of prior works have focused on accelerating NTT,
including GPU-based~\cite{iccis-2015-cuhe, iacr-2018-fvgpu, goey-2020-accelerating} and FPGA-based~\cite{riazi-2020-HEAX, kim-2020-hardware} approaches,
which exploited the algorithmic similarity between DFT and NTT and utilized the same FFT (fast Fourier transform)-based optimizations on NTT.

However, these studies did not identify the primary algorithmic differences between NTT and DFT, which profoundly affect how to implement these algorithms that are tailored to modern GPUs.
Moreover, they focused on implementations with small parameter sets lacking the capability of bootstrapping~\cite{cheon2018bootstrapping}, 
which is highly desirable with HE, which runs sophisticated, real-world applications, 
or they applied standard FFT-based optimizations without providing insights into GPU resource utilization.

The main difference with regard to performing NTT and DFT on GPUs is that the size of the precomputed tables of NTT is far larger than that of DFT.
Compared to floating-point multiplications for DFT, the integer modular operation required with NTT is computationally expensive on modern GPUs.
Typical NTT implementations~\cite{aguilar-2016-nfllib,icfcds-2017-seal} use Shoup’s modular multiplication (\emph{modmul})~\cite{jsc-2014-harvey} 
to reduce the high cost of the modular multiplication on GPUs, requiring the same number of precomputed values as the number of twiddle factors.

The input data of NTT for an HE operation is a sequence of residues with the prime modulus used in CRT.
The number of primes ($np$) used as the moduli amounts to several dozens.
Because the root of unity differs for each prime modulus, there is an $np$-fold increase in the size of the twiddle factors required to perform $np$ independent NTT operations (e.g., in the $N$-point NTT case, a $2 \times N \times np$-fold increase).

Moreover, the size of the precomputed tables increases significantly as the values of $N$ and $np$ increase to support the bootstrapping operation in HE.
Bootstrapping, an operation to reset the noise of encrypted data accumulated in the course of HE operations, make more operations applicable to ciphertexts without requiring decryption.
Without bootstrapping,
the number of operations in a ciphertext domain is limited;  therefore, HE requires intermittent bootstrapping operations to compute encrypted data continually.
However, as bootstrapping itself contains many HE operations, the values of $N$ and $np$ increase, requiring the size of the precomputed tables to surpass several dozens of megabytes.
Because they do not fit into the on-chip memory of modern GPUs~\cite{tesla-2018-v100}, NTT suffers from a main-memory bandwidth bottleneck.

In this paper, we distinguish the algorithmic characteristics of NTT against DFT
and dissect the effects of this difference when performing NTT on modern GPUs. 
Furthermore, we perform a comprehensive study of the trade-off and performance issues when applying the FFT-based optimizations and algorithms to NTT. 
Specifically, we conduct a performance analysis of the Cooley-Tukey and the Stockham algorithms with various radix values 
and hardware-specific features of GPUs, such as shared memory~\cite{nvidia-2019-ptx}.

By means of this performance analysis, we identify the main-memory bandwidth bottleneck induced by the algorithmic characteristics of NTT
and suggest NTT-specific on-the-fly root generation. 
We compute some part of the twiddle factors on the fly with a novel on-the-fly twiddling (OT) technique to reduce the cost of modular multiplication. 
OT reduces main-memory access and relieves the memory boundedness of NTT.

In summary, we make the following key contributions:

\begin{itemize}
  \item We analyze the primary difference between DFT and NTT in the HE application domain performed on modern GPUs 
  in terms of the computational requirements and performance characteristics.
  \item We conduct a comprehensive study of design space for various optimizations commonly applicable to NTT and DFT. 
  \item Through the comprehensive analysis, we identify that NTT is limited by the main-memory bandwidth after a series of optimizations are applied
  and propose a new on-the-fly root generation technique which results in an additional speedup of 9.3\% on average.
\end{itemize}

\section{An Overview of GPUs}
\label{sec:background}

\begin{table*}[tb!]
  \centering
  \renewcommand{\arraystretch}{1.2}
  \caption{GPU specific terminology, acronym, and description.  NVIDIA Titan V~\cite{tesla-2018-v100} was used for evaluation.}
  \label{table:acronyms}
  \begin{tabular}{lcp{4.5in}l}
    \toprule
    \textbf{Terminology} & \textbf{Acronym} & \textbf{Description}\\
    \midrule
    Streaming Multiprocessor & SM & A unit of computing cores running the same GPU kernel.\\ 
    Thread block & & A group of threads allocated to an SM.\\
    Grid & & A group of blocks launched per GPU kernel.\\
    Warp & & A group of 32 threads in a block, which are scheduled together.\\
    Global memory & GMEM & Main memory space with high latency and low throughput. ($\leq$ 24GB).\\
    Local memory & LMEM & Memory space where per-thread values are automatically stored by the compiler when register spill occurs; LMEM has the same latency and throughput as GMEM.\\
    Texture memory & TMEM & Read-only memory space optimized for accesses with a 2D spatial locality. ($\leq$ 24GB).\\
    Constant memory & CMEM & Read-only memory space whose performance is tailored to the cases when the threads in a warp access the same value ($\leq$ 64KB).\\
    Shared memory & SMEM & Per-block memory space used as a scratchpad ($\leq$ 128KB per SM).\\
    \bottomrule
  \end{tabular}
\end{table*}
This section explains the pertinent details of the organization and operations of modern general-purpose GPUs (GPGPUs).
Table~\ref{table:acronyms} lists NVIDIA GPGPU-specific terms that are frequently used in this paper, and the typical sizes of the logical memory space in modern GPUs~\cite{nvidia-2019-ptx}.
A GPU consists of a number of scalar, in-order processors that exploit massive thread-level parallelism. These scalar processors are grouped to form Streaming Multiprocessors (SMs). An SM, with its own register file, caches, and control units, is an entity that distributes and schedules threads to scalar processors.

GPUs provide different types of logical memory spaces, such as global memory (GMEM), texture memory (TMEM), constant memory (CMEM), and shared memory (SMEM). 
Each memory space has its own features.
GMEM generally plays the main memory role, being large but slow, especially for data without sufficient locality. 
CMEM is a read-only memory space that performs well when multiple threads access the same data.
TMEM targets memory accesses exhibiting a 2D spatial locality. 
SMEM is a type of scratchpad memory shared by a group of threads called a thread block. 
It can read and write data with latency values similar to those of the L1 cache in modern GPUs~\cite{jia-2018-dissecting}.

Registers are the fastest memory in GPUs. 
The number of register entries required for a GPU kernel is calculated during compilation. 
Because the capacity of register files is limited ($\le$ 256KB per SM), if the kernel requires too many registers, a register spill occurs to GMEM, and the spilled memory space is called local memory (LMEM). 
Because LMEM has the same latency and throughput as GMEM, it can degrade the performance of memory-bounded applications.

On a GPU, the latency of each instruction is hidden by executing ready instructions from another thread. 
However, because resources such as SMEM and registers are scarce, if each thread occupies a large amount of these resources, fewer threads can run simultaneously on an SM; 
the ratio of the number of concurrently running threads over the maximum of a machine is called the \emph{occupancy} rate. 
When the kernel occupancy becomes too low to hide the stalls of the instructions being executed, the performance of the kernel decreases.

A GPU kernel is launched from the host side. 
The launched kernel groups several threads to form a thread block; a group of blocks forms a grid. 
Each thread block is allocated to one SM. 
Modern GPUs provide synchronization between threads in a thread block such that those threads can share SMEM without data races~\cite{nvidia-2019-ptx}.

SM schedules a group of threads called a \emph{warp}. 
Each warp consists of 32 consecutive threads. 
When the threads of a warp request data, if the addresses of the requested data are continuous, the requests are merged into fewer memory transactions, 
each of which is 32 bytes.
This process it is referred to as \emph{memory coalescing}. Therefore, depending on the data access pattern, 
different numbers of actual memory transactions are required for a given amount of requested data. 
If the addresses of the requested data within a warp are not sufficiently continuous, 
the number of memory transactions required is bloated, 
leading to poor performance~\cite{bell-2009-implementing}. 
Therefore, it is critical to program the memory access pattern carefully.

We utilize a NVIDIA Titan V GPU~\cite{tesla-2018-v100} as our target architecture and use it for all of the experiments throughout this paper.
It consists of 80 SMs, each having 64 computing cores.

\section{An Overview of the Key Algorithms}
\label{key_algo}

This section explains the key algorithm, Number Theoretic Transform (NTT), and how NTT is used for HE.
Then, we describe a basic implementation of NTT and the method used to apply FFT algorithms to NTT.

\subsection{Number Theoretic Transform (NTT) and Modular Polynomial Multiplication}
\emph{NTT} is an algorithm that is a specialized form of the Discrete Fourier Transform (DFT) for a finite field of integers.
DFT chooses $e^{-2\pi j/N}$ as the primitive $N_{th}$ root of unity and generates twiddle factors from the exponentials of the root of unity. 
NTT chooses $\psi$ as the primitive $N_{th}$ root of unity where $\psi^N \equiv 1\;\mathbf{mod}\;{p}$ for a given $N$ and a prime $p$ such that $p = kN + 1$,
generating twiddle factors from the exponentials of the primitive $N_{th}$ root of unity. 
FFT multiplies complex numbers whereas NTT performs integer modular multiplication. 
The overall NTT algorithm computes the following:
$X_k = (\sum_{n=0}^{N-1}{x_{n}\psi_{N,p}^{n(k)}})\;\mathbf{mod}\;{p}$ where $0 \leq k < N$,
$\psi_{N,p}$ is the primitive $N_{th}$ root of unity of NTT for $Z_p$,
$x_n$ is an input polynomial coefficient indexed by $n$, and
$X_k$ is an output polynomial coefficient indexed by $k$.

We can exploit NTT to undertake the multiplication in a polynomial ring $Z[X]/(X^N+1)$.
The coefficients of the output of multiplication of two polynomials in the polynomial ring 
are equal to the output of negacyclic convolution of the coefficients of the two polynomials.
For example, when
  $A(X) = \Sigma_{k=0}^{N-1}a_k X^k \in Z[X]/(X^N+1)$ 
  and $B(X) = \Sigma_{k=0}^{N-1}b_k X^k \in Z[X]/(X^N+1)$,
\begin{equation*}
  C(X) = A(X) \cdot B(X)\;\mathbf{mod}\;{(X^N+1)}= \Sigma_{k=0}^{N-1}c_k X^k
\end{equation*}
  where $c_k = \sum _{i=0}^{k}a_i b_{k-i} - \sum _{i={k+1}}^{N-1}a_i{b_{N+k-i}}$.

This convolution operation is called negacyclic convolution because the sign of the
second term is negative.
At the same time, as in DFT, element-wise multiplication in the NTT domain, followed by an iNTT, is exploited to perform negacyclic convolution as shown in the following relationship~\cite{pop-2012-towards}:
\begin{equation*}
  \mathbf{c} = \mathbf{\Psi}^{-1} \odot iNTT(NTT(\mathbf{\bar{a}}) \odot NTT(\mathbf{\bar{b}}))
\end{equation*}
where the operator $\odot$ denotes the element-wise multiplication of vectors, $\mathbf{\bar{a}}$, $\mathbf{\bar{b}}$, and $\mathbf{c}$ are the vectors of the coefficients of $A(\psi_{2N,p} \cdot X)$, $B(\psi_{2N,p} \cdot X)$, and $C(X)$, respectively; and $\mathbf{\Psi}^{-1} = (1, \psi_{2N,p}^{-1}, \psi_{2N,p}^{-2}, \ldots, \psi_{2N,p}^{N-1})$.
We use bold letters to indicate vectors.
We can merge the term $\psi_{2N, p}$, which appears in $A(\psi_{2N, p} \cdot X)$ and $B(\psi_{2N, p} \cdot X)$, with the following NTT to reduce the number of computations. For example, the output of $NTT(\mathbf{\bar{a}})$, $\mathbf{A} = (A_0, A_1, \ldots, A_{N-1})$, is computed as follows~\cite{roy-2014-compact}:
\begin{equation*}
\begin{split}
  A_k & =\sum_{n=0}^{N-1}{(a_{n}\psi_{2N,p}^n)}\psi_{N,p}^{n \cdot k} \\
  & =\sum_{n=0}^{N-1}{(a_{n}\psi_{2N,p}^n)}\psi_{2N,p}^{2n \cdot k}=\sum_{n=0}^{N-1}{a_{n}\psi_{2N,p}^{n(2k+1)}}
\end{split}
\end{equation*}

Similarly, the element-wise multiplication of $\mathbf{\Psi}^{-1}$ can be merged into the iNTT~\cite{poppelmann-2015-high}. 
%
By adapting FFT algorithms~\cite{mc-1965-ntt,cochran-1967-fast} for use in the merged NTT and iNTT, which will be explained in a later section,
the computational complexity of the multiplication between two polynomials in the polynomial ring $Z[X] / (X^N+1)$ is reduced from $O(N^2)$ of a na\"ive convolution to $O(NlogN)$.

\subsection{NTT in Homomorphic Encryption}

Encrypted data in HE is stored in a cyclotomic polynomial ring $\in Z_Q[X]/(X^N+1)$, which has a ciphertext modulus $Q$ with a big-integer size ($Q \gg 2^{64}$). 
Because multiplying encrypted messages requires modular polynomial multiplications in the ring
and the value of $N$ is typically large (e.g., $2^{13}$ to $2^{17}$), HE adopts NTT to perform the modular polynomial multiplications.
Because the coefficients of the polynomials are big integers $\mathbf{mod}\;Q$, 
multiplying these coefficients is computationally expensive. 
HE schemes~\cite{bajard-2016-full,cheon-2018-full,han-2020-better} reduce the cost of such multi-precision multiplication by employing the Chinese remainder theorem (CRT), 
which transforms big integer coefficients to residue number system (RNS) representations. 
CRT states that given $np$ coprimes $(p_1,p_2,\ldots,p_{np}) \ni\:s.t.\;\Pi{p_i} \ge Q$, an arbitrary integer smaller than $Q$ is uniquely represented by remainders $(r_1, r_2, …, r_{np})$ whose moduli are the $np$ coprimes.

Typical HE schemes exploit CRT with $np$ primes, each being congruent to 1 $\mathbf{mod}\:N$.
Therefore, a polynomial in $Z_Q[X]/(X^N+1)$ is divided into $np$ polynomials where the
$i_{th}$ polynomial is in $Z_{p_i}[X]/(X^N+1)$.
Finally, the multiplication operations in $Z_Q[X]/(X^N+1)$, which are multi-precision
operations, become element-wise modular multiplications in $Z_{p_i}[X]/(X^N+1)$ for $i$ in $[1, np]$.

Putting it all together, for a polynomial having a degree up to $N$ and represented in an
RNS domain with $np$ primes, $np$ $N$-point NTTs are required to perform multiplication
with another polynomial.
Although affected by specific HE schemes and parameter settings,
typical values of $N$ and $np$
are from $2^{14}$ to $2^{17}$ for $N$ and range up to several dozens for $np$.
Therefore, the size of a polynomial reaches dozens of megabytes.

\begin{algorithm}[tb!]
\caption{Cooley-Tukey NTT}
\small
\label{algorithm:cooley-tukey ntt}
\begin{algorithmic}[1]
  \REQUIRE{A vector $\mathbf{a}=(a[0], a[1], \ldots,a[N-1])$ and $a[i]_{0 \le i < N}$ = the $i_{th}$ coefficient of $A(x)\;\mathbf{mod}\;p$, where 
   $A(X) = \Sigma_{k=0}^{N-1}a_k X^k \in Z_Q[X]/(X^N+1)$ and $p$ is a prime such that $p \equiv 1 \;\mathbf{mod}\; 2N$.
  A vector $\mathbf{\Psi}=(\Psi[0],\Psi[1],\ldots,\Psi[N-1])$ and $\Psi[i]_{0 \le i < N}=\psi_{2N,p}^{bit-reverse(i)}$.} 
  \ENSURE{$\mathbf{a} \leftarrow NTT(\mathbf{a})$ in a bit-reverse order.} 
  \State {$t =  N\mathrel{/}2$}
  \FOR{$(m = 1; \; m < N; \; m \gets m \cdot 2)$}
    \FOR{$(j = 0; \; j < m; \; j \gets j+1)$}
    \FOR{$(k = j \cdot 2t; \; k < j \cdot 2t + t; \; k \gets k+1)$}
      \State{$\mathbf{Butterfly}(a[k],a[k+t],p,\Psi[m+j])$}
    \ENDFOR
    \ENDFOR
	  \State {$t = t \mathrel{/} 2$}
	\ENDFOR
\end{algorithmic}
\end{algorithm}

\subsection{Basic Implementation of NTT with Cooley-Tukey and Stockham algorithm}
The Cooley-Tukey~\cite{mc-1965-ntt} and the Stockham~\cite{cochran-1967-fast} algorithms are the two most popular FFT algorithms.
They reduce the computation complexity of a na\"ive DFT from $O(N^2)$ to $O(Nlog{N})$.

Prior works~\cite{iccis-2015-cuhe,icfcds-2017-seal,iacr-2018-fvgpu} that accelerate NTT exploited the Cooley-Tukey algorithm. 
The algorithm recursively divides an $N$-point DFT into $k$ interleaved $N/k$-point NTTs.
Depending on the divisor $k$, the algorithm is called the radix-$k$ NTT, and each division is called \emph{stage}; the number of stages becomes $log_k{(N)}$. 
Algo.~\ref{algorithm:cooley-tukey ntt} shows the pseudo-code for a na\"ive radix-2 NTT. 
At each stage (variable $m$ in Algo.~\ref{algorithm:cooley-tukey ntt}), multiple butterfly operations (Algo.~\ref{algorithm:butt}) are performed. 
The twiddle factors used during the butterfly operations are stored in a precomputed table, \emph{$\mathbf{\Psi}$}. The number of twiddle factors required doubles at each stage.

\begin{algorithm}[tb!]
\caption{Butterfly operation}
\small
\label{algorithm:butt}
\begin{algorithmic}[1]
  \REQUIRE{$0 \le A < 4p, 0 \le B < 4p, p, 0 \le \Psi < p$}
  \ENSURE{$A,B$}
  \State {$\bar{B} = (B \times  \Psi)\;\mathbf{mod} \; p$}
  \State {$B = A \mathrel{-} \bar{B}$}
  \State {$A = A \mathrel{+} \bar{B}$}
\end{algorithmic}
\end{algorithm}
Merging the powers of the $\psi_{2N,p}$ term in the negacyclic convolution using NTT/iNTT with the powers of $\psi_{N,p}$ in NTT is also possible in the Cooley-Tukey algorithm~\cite{roy-2014-compact}. 
The difference from NTT without negacyclic convolution is that 
it uses the powers of the $2N_{th}$ root of unity for twiddle factors ($\mathbf{\Psi}$ in Algo.~\ref{algorithm:cooley-tukey ntt}), 
which are stored in a bit-reverse order:

\begin{itemize}
  \item[] $\Psi[j] = \psi^{bit-reverse(j)}_{2N,p}$
\end{itemize}

Because the Cooley-Tukey algorithm produces output data in a bit-reverse order, the input data to the algorithm requires bit-reversal permutation (bit-reversing) prior to or after NTT to reorder the output values.

The Stockham algorithm~\cite{cochran-1967-fast} recursively divides an $N$-point NTT into $N/k$-point NTTs.
As opposed to the Cooley-Tukey algorithm, Stockham does not need any extra permutation to obtain an aligned output; instead, Stockham stores the permuted output at each stage.
The pseudo-code of the na\"ive radix-2 Stockham algorithm is shown in Algo.~\ref{algorithm:Stockham ntt}.
\begin{algorithm}[tb!]
\caption{Stockham NTT}
\small
\label{algorithm:Stockham ntt}
\begin{algorithmic}[1]
  \REQUIRE{A vector $\mathbf{a}=(a[0], a[1], \ldots,a[N-1])$ and $a[i]_{0 \le i < N}$ = the $i_{th}$ coefficient of $A(x)\;\mathbf{mod}\;p$, where 
   $A(X) = \Sigma_{k=0}^{N-1}a_k X^k \in Z_Q[X]/(X^N+1)$ and $p$ is a prime such that $p \equiv 1 \;\mathbf{mod}\; 2N$.
  A vector $\mathbf{\Psi}=(\Psi[0],\Psi[1],\ldots,\Psi[N-1])$ and $\Psi[i]_{0 \le i < N}=\psi_{2N,p}^{bit-reverse(i)}$. 
  A vector $\mathbf{A}=(A[0], A[1],\dots,A[N-1])$ and $A[i]_{0 \le i < N}$ is initialized to zero.}
  \ENSURE{$\mathbf{A} \leftarrow NTT(\mathbf{a})$.} 
  \State{$t =  N\mathrel{/}2$}
  \FOR{$(m = 1; \; m < N; \; m \gets m \times 2)$}
    \FOR{$(j = 0; \; j < m; \; j \gets j+1)$}
    \State {$b = j \cdot N/m/2$}
    \FOR{$(k = 0; \; k < N/2/m; \; k \gets k+1)$}
      \State{$\mathbf{Butterfly}(a[b+k],a[b+k+N/2],p,\Psi[m+k])$}
      \State{$A[2 \cdot m \cdot j+k] = a[b+k]$}
      \State{$A[2 \cdot m \cdot j+k+m)] = a[b+k+N/2]$}
    \ENDFOR
    \ENDFOR
    \FOR{$(j = 0;\; j< N;\; j\gets j+1)$} 
      \State{$a[j] = A[j]$}
    \ENDFOR
	  \State{$t = t \mathrel{/} 2$}
	\ENDFOR
\end{algorithmic}
\end{algorithm}

In the Stockham algorithm, the result is produced in order without bit-reversing.
However, as the addresses of the input and output values of a butterfly differ, the Stockham algorithm must be performed in an out-of-place manner to prevent data races~\cite{govindarajuo-2008-high}.

\section{Comparing NTT with DFT}
\label{sec:DFT-NTT Comparison}
In this section, we assess the fundamental differences and possible design choices when applying the basic FFT algorithm to NTT and DFT running on GPUs.

\textbf{Cooley-Tukey vs. Stockham:}
Prior works~\cite{govindarajuo-2008-high, lloyd-2008-fast} accelerating DFT on GPUs mostly use the Stockham algorithm 
rather than the Cooley-Tukey algorithm, as Cooley-Tukey requires an extra bit-reversing step, 
which is less friendly to memory coalescing.
Therefore, it requires redundant memory accesses, degrading the performance.

In contrast, NTT used for HE operations does not require bit-reversing in nature. 
The output of NTT in HE only performs integer modular multiplication, in an element-wise manner, 
with another output of NTT; whether the output values are permuted or not is irrelevant.

Moreover, the Stockham algorithm requires out-of-place computation~\cite{govindarajuo-2008-high}.
Therefore, the working set increases and the caches behave less effectively as the data sizes of the DFTs and the NTTs increase. 
\cite{govindarajuo-2008-high} proposed an in-place Stockham implementation by exploiting shared memory.
However, this method also requires extra transpose operations on the input data and therefore does not perform better than the Cooley-Tukey algorithm without bit-reversing. 
Therefore, we choose the Cooley-Tukey algorithm and use it throughout this paper.

\textbf{32b vs. 64b word size:}
The input of DFT is a sequence of complex numbers whose real and imaginary parts are single (32b) 
or double (64b) floating-point numbers depending on the application-specific precision requirements. 
In contrast, the word size of the input of NTT depends on the sizes of the different prime moduli. 
Each prime modulus is typically a single (32b) or double (64b) integer word. 
The choice of the moduli and their size depends on each HE scheme.

One notable difference in NTT is that the choice of moduli size affects the size of the entire workload (i.e., the total number of butterfly operations).
For example, suppose that $Q=2^{1200}$. 
Then 40 primes are needed with 30-bit primes (requiring single-word operations), whereas only 20 primes are adequate with 60-bit primes (requiring double-word operations).

This trade-off between the operational complexity and the workload size has a negligible effect on the performance of NTT.
We compared the performance between 32b- and 64b-operation-based NTTs after applying the optimizations 
explained in the later sections, showing that the performance difference is approximately 5\% with the parameters of $N=2^{17}$ and $Q=2^{1200}$.
We choose 64b as the word size and the prime numbers between $2^{59}$ and $2^{60}$.

\begin{figure}[!tb]
  \center
  \includegraphics[height=0.7in]{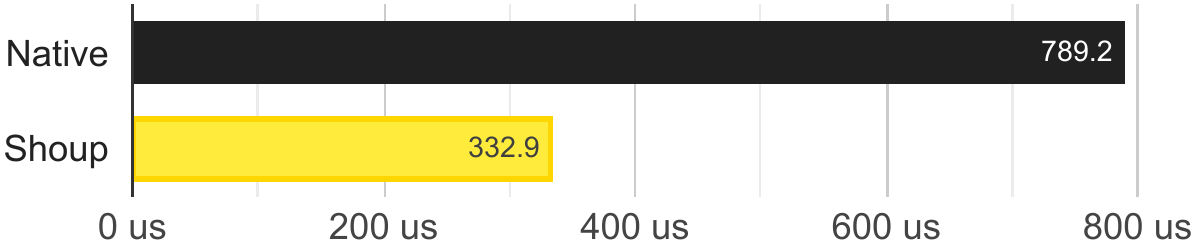}
  \vspace{0.05in}
  \caption{
    NTT performance with Shoup's modmul (Shoup) and native modulo operation provided by CUDA (Native).
    We use parameters ($N$, $np$) as ($2^{17}$, 45).
  }
  \label{fig:native_shoup}
  \vspace{-0.05in}
\end{figure}

\textbf{Floating-point multiplication vs. integer modular multiplication:} 
DFT performs floating-point multiplications. 
On the other hand, NTT performs integer multiplications with modular reductions.
When the word size is 32b, modern GPUs support native modulo operations (i.e., a double-word modulo a single-word). 
However, the native modulo operation is significantly slower in terms of the latency; our experiments with a simple benchmark show that performing a 64b integer modulo a 32b integer is compiled to 68 machine instructions~\cite{nvidia-2020-sass} with latency of around 500 cycles.

Barrett reduction~\cite{barrett-1986-implementing} and Shoup’s modmul~\cite{shoup-2001-ntl} (Algo.~\ref{algo:shoup}) are two ways to mitigate the computational cost of integer modular multiplication.
Although both require pre-computed data to perform modular reduction and utilize more memory bandwidth, they still outperform the native modular reduction.
Figure~\ref{fig:native_shoup} shows that NTT with Shoup’s modmul has a speedup of 2.4 over the one with modular multiplication using the native modulo operation.

\begin{algorithm}[tb!]
\caption{Shoup's modular multiplication (modmul)}
\small
\label{algo:shoup}
\begin{algorithmic}[1]
  \REQUIRE{$p < \beta / 4,0 \le b < 4p, 0 \le w < p$}
  \REQUIRE{$\bar{w} \gets \{w\times\beta/p\},\beta \gets 2^{32} \; or \; 2^{64}$}
  \ENSURE{$r = \mathbf{mod}(b \times w, p)$}
  \State{$q=(b\times \bar{w}) \ll log(\beta)$}
  \State{$r=b \times w - q \times p$}
  \IF{$r > p$}
    \State{$r=r -  p$}
  \ENDIF
\end{algorithmic}
\end{algorithm}

\textbf{Precomputed table size with batching:}
Multiple $N$-point DFTs can be executed together, forming a batch (batching)~\cite{govindarajuo-2008-high}.
With batching, performing DFTs still requires $N$ twiddle factors, which is identical to performing a single DFT. 
In contrast, the number of twiddle factors required by NTT is proportional to the batch size.
Many prior works on DFT~\cite{kopcke-2019-generating, chang-2016-accelerating, nukada-2008-bandwidth} pre-computed twiddle factors because cosine and sine operations supported by GPUs have low throughputs. 
When batching multiple DFTs, the cost (the size per DFT) of the precomputed table is amortized because the $N$ twiddle factors are shared and reused within a batch: the same primitive $N_{th}$ root of unity is used for any $N$-point DFT.
On the other hand, the primitive $N_{th}$ root of unity differs for each NTT with different primes, increasing the number of twiddle factors needed by $np$ times when batching $np$ independent NTT.
Moreover, Shoup’s modmul requires extra precomputed values ($\bar{w}$ in Algo.~\ref{algo:shoup}), doubling the sizes of the precomputed tables.

\section{Optimizations that are Commonly Applicable to NTT and DFT}
\label{sec:contribution-2}
In this section, we introduce FFT-based optimization techniques that are applicable to both DFT and NTT.

\textbf{Batching FFT with various batch sizes:} 
Prior works~\cite{stvrelak-2018-performance,govindarajuo-2008-high} accelerated DFT by batching multiple independent DFTs with the same size.
HE can also exploit the batching technique because it performs $np$ independent NTTs, each having a different prime modulus.

\textbf{Register-based high-radix implementation:} 
Each GPU thread in Algo.~\ref{algorithm:cooley-tukey ntt} accesses GMEM twice to retrieve two input operands per butterfly operation. 
By using a higher radix (e.g., $2^k$), a thread takes $2^k$ input data at a time, stores them into registers, 
and performs $2^k$-point NTT.
This approach reduces access to GMEM by $k$ times.

\begin{figure}[!tb]
  \center
  \includegraphics[height=2.2in]{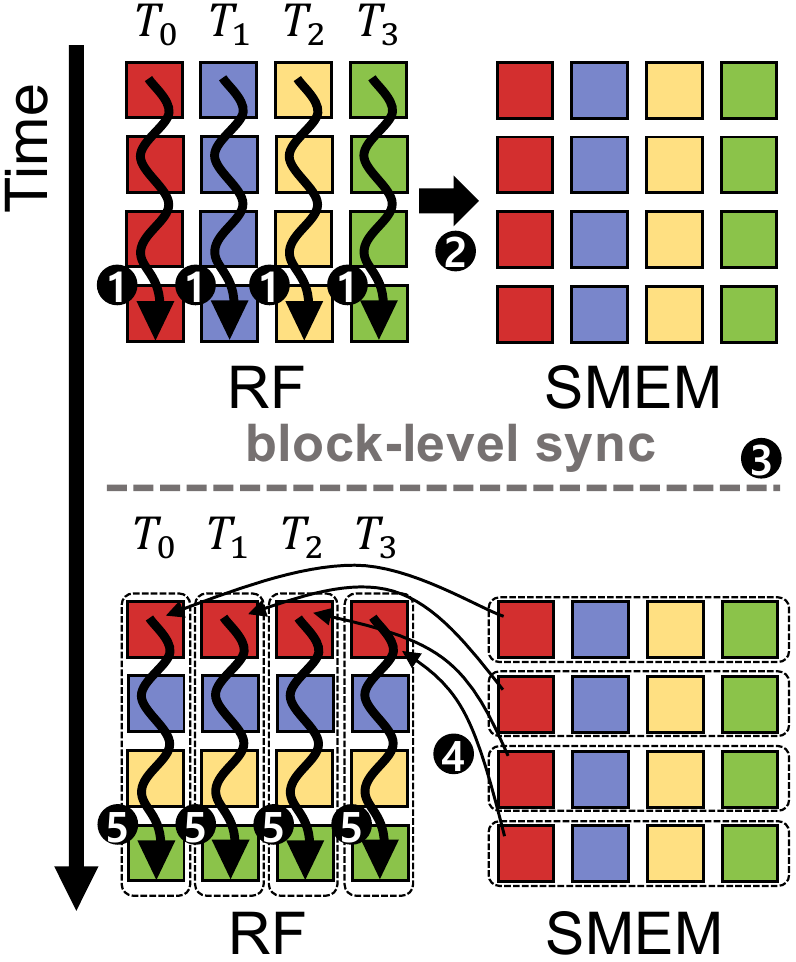}
  \vspace{0.0in}
  \caption{
Radix-4 $\times$ 4 NTT with shared memory implementation. The 16 input data are shown as squares. {\encircled{1}} Each thread of $T_0$, $T_1$, $T_2$, and $T_3$ performs 4-point NTT with data stored in a register file (RF), and {\encircled{2}} stores the outputs to SMEM. Then, there comes a {\encircled{3}} block-level synchronization, followed by {\encircled{4}} loading the data from SMEM to RF in a transposed manner. {\encircled{5}} Finally, each thread performs 4 per-thread NTT.
  }
  \label{fig:smem_impl}
  \vspace{-0.05in}
\end{figure}

\textbf{Shared memory (SMEM) implementation~\cite{govindarajuo-2008-high}:}
Because the number of registers in an SM is limited, we cannot increase radix indefinitely. 
SMEM implementation complements the register-based high-radix implementation by using fewer registers per thread.
During the implementation of radix-$r1 \times r2$ NTT, a thread performs $r1$-point NTT, 
stores the output in SMEM,
synchronizes with other threads in the same block to avoid a data race, and then performs $r2$-point NTT whose input values are loaded from SMEM in a transposed manner.
During the implementation of SMEM, the size of NTT performed by a single kernel is different from that by a single thread. 
For distinction, we refer to the size of NTT performed by a single kernel as radix 
and the size of NTT performed by a single thread as \emph{per-thread NTT}.
Figure~\ref{fig:smem_impl} shows an example of radix-4 $\times$ 4.

This shared memory implementation reduces the register pressure from $O(R)$ to $O(\sqrt{R})$ when using radix-R compared to the high-radix implementation.
Although SMEM is not as fast as the registers which can fetch data in a single cycle, it has latency and throughput values similar to those of the L1 cache.

\textbf{Caching twiddle factors:}
Previous studies~\cite{zhang-2017-gpu,ulu-2020-high,nukada-2008-bandwidth} have attempted to reduce GMEM accesses by storing twiddle factors in read-only memory spaces, TMEM and CMEM. 
Because CMEM is small (64KB~\cite{tesla-2018-v100}), it is difficult to contain all of the twiddle factors of NTT for the HE parameter sets whose ranges are specified in Section~\ref{sec:DFT-NTT Comparison}. 
In contrast, TMEM is as large as GMEM and thereby can be used for NTT.

\section{An Analysis of the Commonly Applicable Optimizations} 
\label{sec:contribution-3}
We apply the aforementioned FFT-based optimizations described in Section~\ref{sec:contribution-2} to NTT
and analyze the computational characteristics and performance of NTT 
with an emphasis on highlighting the key differences from the FFT implementation.

\subsection{Batching NTT with various batch sizes}
Batching increases the throughput of DFT by increasing GPU utilization.
Only executing a single DFT at a time does not fully exploit the hundreds of thousands of threads provided by a modern GPU.

\begin{figure}[!tb]
  \center
  \subfloat[Performance of NTT]{\includegraphics[width=0.48\columnwidth]{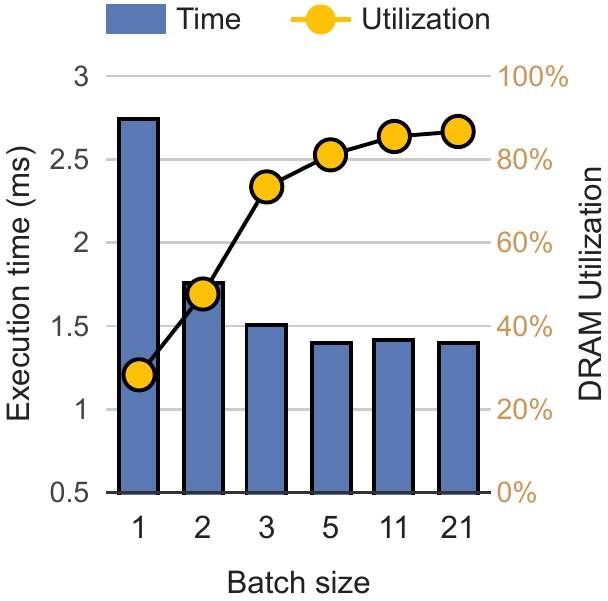}}
  \hspace{0.02in}
  \subfloat[Performance of DFT]{\includegraphics[width=0.48\columnwidth]{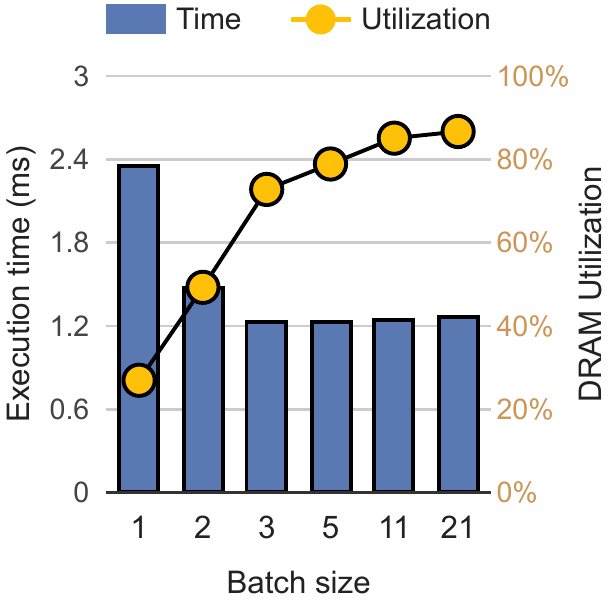}}
  \caption{
The execution time of radix-2 implementation of (a) NTT and (b) DFT with various batch sizes. We use (N, np) as ($2^{17}$, 21).
  }
  \label{fig:batching}
  \vspace{-0.05in}
\end{figure}

First, We point out that batching is effective in NTT. 
Figure~\ref{fig:batching}(a) shows the performance of a radix-2 implementation of $2^{17}$-point NTT with various batch sizes. 
As the batch size increases, the per-NTT performance initially increases and becomes saturated past a batch size of 5. 
When comparing the batch sizes of 1 and 21, the per-NTT execution times are 2751.5 us and 1426.4 us,
respectively, corresponding to a 1.92$\times$ speedup. 
With a large enough batch size, NTT becomes limited by the main-memory bandwidth; for example, with a batch size of 21, the evaluated GPU achieves 86.7\% of its peak main-memory bandwidth (564.4 GB/s).
DFT also shows a similar trend; 
the performance improves as the batch size increases. 
Figure~\ref{fig:batching}(b) shows the performance of our custom radix-2 FFT implementation without bit-reversing.
By batching 21 DFTs together, we obtain a speedup of 1.84$\times$, saturating the main-memory bandwidth up to 86.7\%.  

\subsection{Register-based high radix implementation} 

\begin{figure*}[!tb]
  \center
  \subfloat[High radix performance ($N=2^{16}$)]{\includegraphics[width=2.25in]{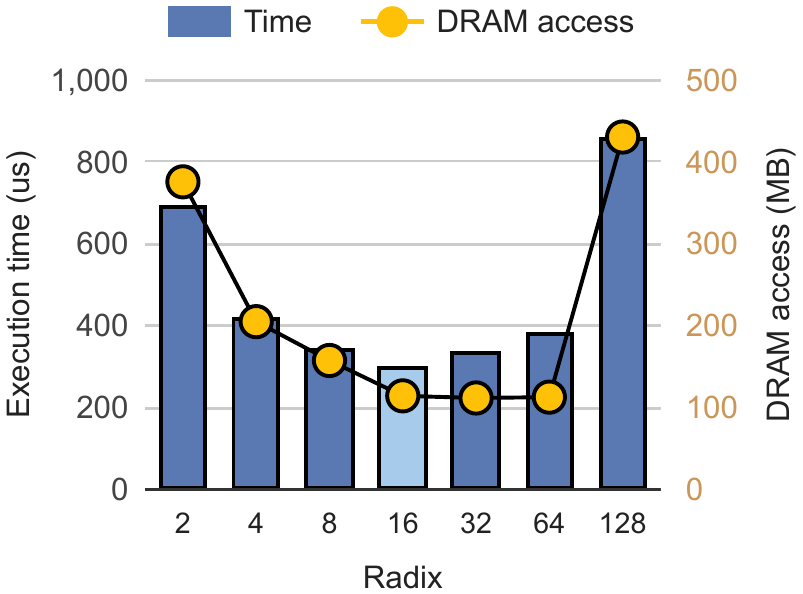}}
  \hspace{0.05in}
  \subfloat[High radix performance ($N=2^{17}$)]{\includegraphics[width=2.25in]{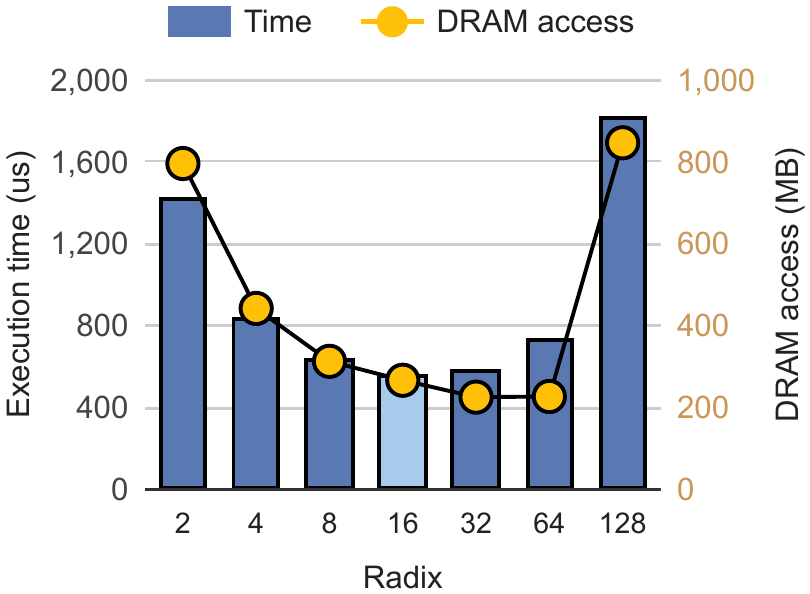}}
  \hspace{0.05in}
  \subfloat[Occupancy \& DRAM throughput ($N=2^{17}$)]{\includegraphics[width=2.25in]{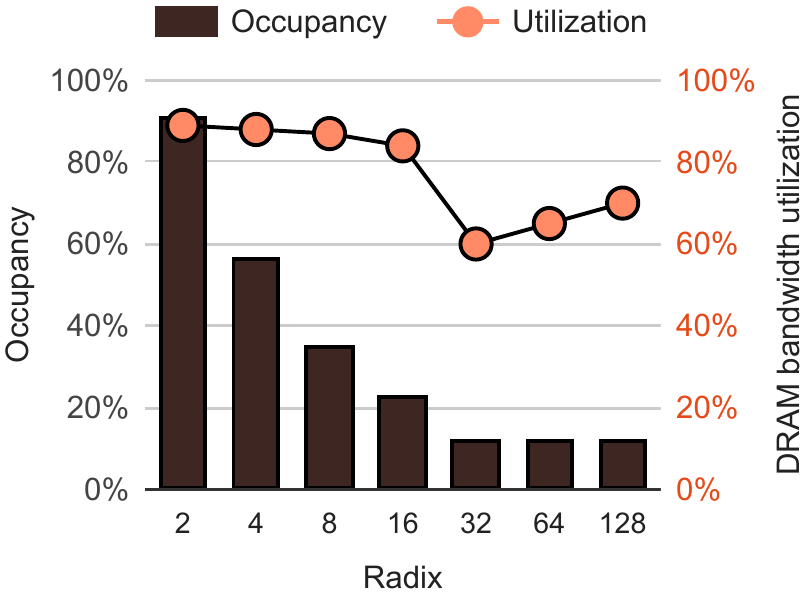}}
  \caption{
  Comparing NTT execution time and the amount of DRAM accesses in different radices when (a) $N=2^{16}$ and (b) $N=2^{17}$, and 
  (c) DRAM bandwidth utilization and occupancy of register-based high radix implementation when using $N=2^{17}$ under the condition of $np=21$.
}
  \label{fig:radix_ntt}
  \vspace{-0.05in}
\end{figure*}

Exploiting registers to enable high-radix implementation significantly improves the performance of NTT 
by reducing the number of DRAM accesses as the radix-2 implementations are limited by the main-memory bandwidth.
However, excessively increasing the radix may degrade the performance 
due to the limited number of registers, as shown in a prior work~\cite{allen-2009-computational} on DFT. 
Figure~\ref{fig:radix_ntt}(a) and Figure~\ref{fig:radix_ntt}(b) correspondingly show the execution time and the number of DRAM accesses while performing NTT with various radices and $N$. 
NTT performs best with radix-16, showing a 2.41 $\times$ speedup on average, compared to the radix-2 cases.

\begin{figure*}[!tb]
  \center
  \subfloat[High radix performance ($N=2^{16}$)]{\includegraphics[width=2.25in]{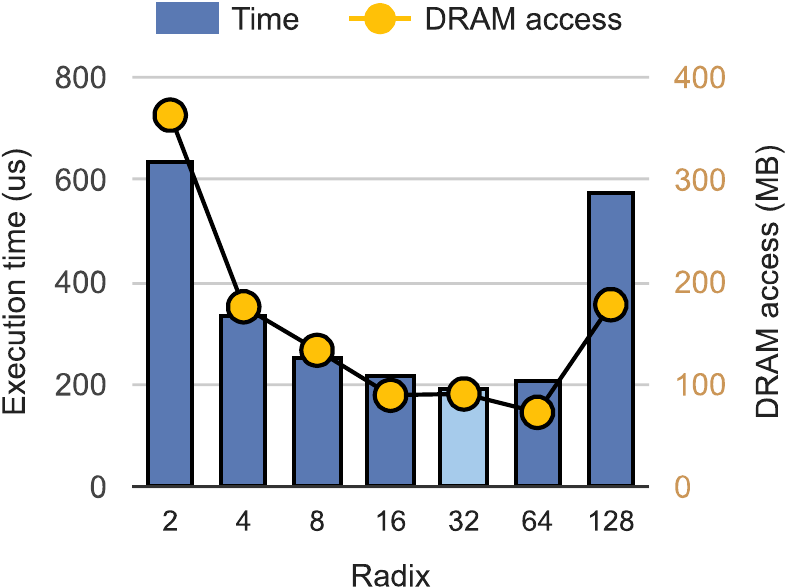}}
  \hspace{0.05in}
  \subfloat[High radix performance ($N=2^{17}$)]{\includegraphics[width=2.25in]{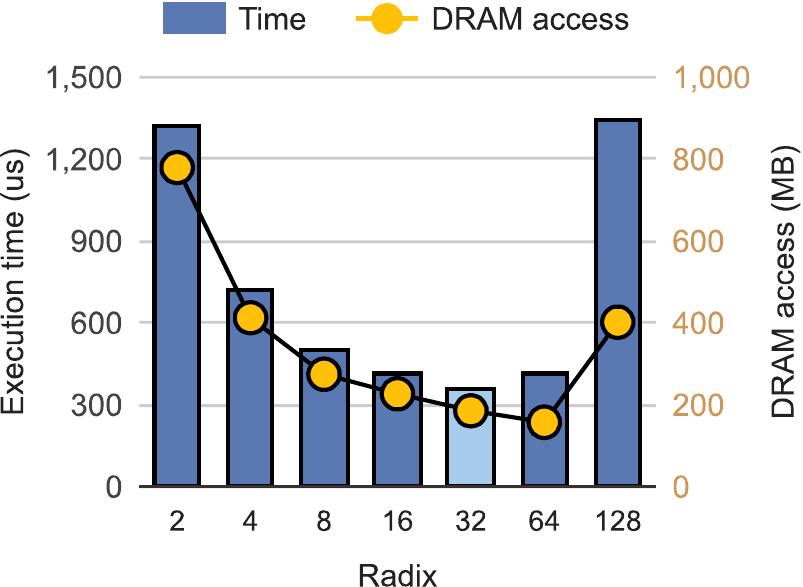}}
  \hspace{0.1in}
  \subfloat[Occupancy \& DRAM throughput ($N=2^{17}$)]{\includegraphics[width=2.25in]{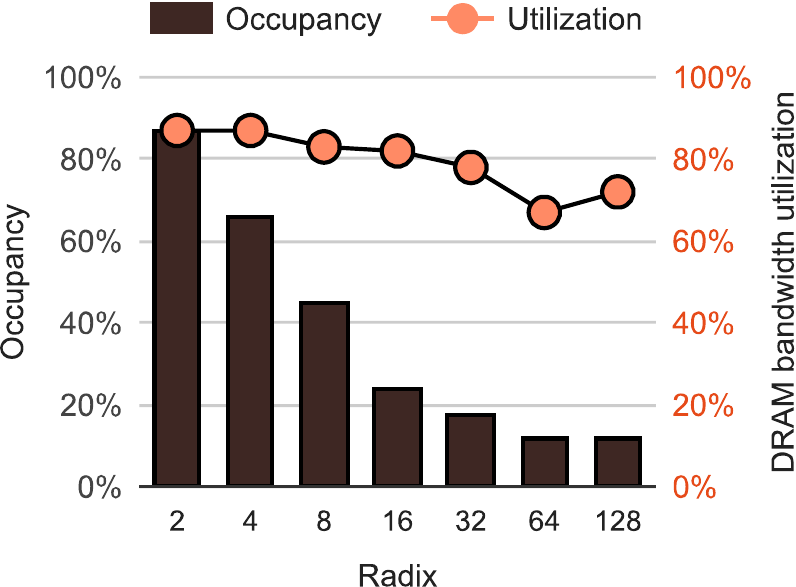}}
  \caption{
  Comparing DFT execution time and the amount of DRAM accesses in different radices when (a) $N=2^{16}$ and (b) $N=2^{17}$, and 
  (c) DRAM bandwidth utilization and occupancy of register-based high radix implementation when using $N=2^{17}$ under the condition of batching 21 sequences of complex numbers.
   }
  \label{fig:radix_dft}
  \vspace{-0.05in}
\end{figure*}

The performance of NTT decreases with radix values higher than 16 due to the register spills and occupancy drops. 
Radix-32 performs worse than radix-16 even if the former has 15.5\% fewer DRAM accesses with $N = 2^{17}$. 
This occurs because the register usage per thread becomes high; 
accordingly, the occupancy becomes too low to utilize the main-memory bandwidth fully
(Figure~\ref{fig:radix_ntt}(c)). 
Therefore, bandwidth utilization falls to 59.9\%. 
The register usage per thread for Radix-64 and radix-128 are too high;
the compiler allocates LMEM instead of registers such that the bandwidth utilization increases while the occupancy remains mostly unchanged.

Each thread of NTT consumes more registers than that of DFT. 
Figure~\ref{fig:radix_ntt}(c) and Figure~\ref{fig:radix_dft}(c) show that the occupancy of NTT is lower by 31.2\% in radix-32, compared to that of DFT. 
The extra amount of register usage of a thread in NTT is for a prime and a precomputed value required in Shoup’s modmul (Algo.~\ref{algo:shoup}). 
The difference in the occupancy causes a difference between the best-performing radix of NTT (radix-16) and that of DFT (radix-32).

\subsection{Shared memory (SMEM) implementation}
SMEM implementation reduces the number of main memory accesses by exploiting SMEM as intermediate storage at each stage. 
This process complements the register-based high-radix implementation with less register usage.
Ideally, to minimize the number of main memory accesses of $N$-point NTT, an SM should accommodate all $N$-point input values such that the SM loads from GMEM only once.
However, the register file and SMEM are too small: the working set of NTT for a single prime ranges in size from 512KB ($N=2^{16}$) to 1MB ($N=2^{17}$), whereas an SM has a register file of 256KB and a SMEM not exceeding 128KB.
Hence, input data cannot be loaded at once, and at least two loads from GMEM are required on a modern GPU.

By exploiting the SMEM implementation, we can load the input data from GMEM only twice, achieving high-performance NTT. 
Our experiments show that the SMEM implementation can increase the radix up to $2^{11}$ without any instances of register spill or a severe occupancy drop that underutilizes memory bandwidth. 
Therefore, two GPU kernels are needed for $N$-point NTT: the first one is for radix-$N1$ NTT, and the second one is for radix-$N2$ NTT, where $N = N1 \times N2$ and both $N1$ and $N2$ are at least 64. We refer to these two kernels as \emph{Kernel-1} and \emph{Kernel-2}, respectively, in the order of kernel execution. 

\textbf{Avoiding uncoalesced memory accesses in Kernel-1:}
According to the definition of the Cooley-Tukey algorithm, for an $N$-point NTT, Kernel-1 initially performs $N2$ $N1$-point NTTs where each thread accesses $N1$ data in a strided fashion (with a large stride value). 
Then, Kernel-2 performs $N1$ $N2$-point NTTs, where each thread accesses $N2$ data continuously located in memory.
A na\"ive implementation of Kernel-1 causes most strided memory accesses to be uncoalesced, leading to wasted memory bandwidth per memory transaction mostly.

\begin{figure}[!tb]
  \center
  \subfloat{\includegraphics[height=1.5in]{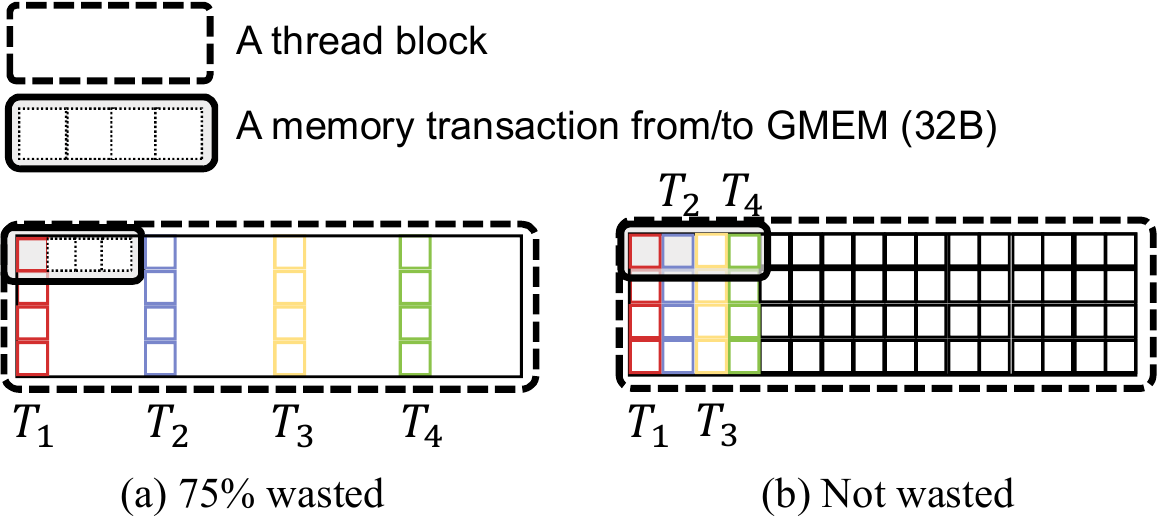}}
  \vspace{0.0in}
  \caption{
  Example of the first radix-16 NTT (Kernel-1) in performing a 64-point NTT.
  Data in GMEM required by threads in a thread block are shown as small squares.
  The first four threads  (T1, T2, T3, and T4) are colored differently.
  (a) In one thread block among the four (the others are not shown), each thread wastes 75\% of a memory transaction.
  (b) By combining the four-thread blocks in (a) into one, and letting each thread perform adjacent 4-point NTT, memory accesses can coalesce.
   }
  \label{fig:coalescing}
  \vspace{-0.05in}
\end{figure}

Prior work on DFT~\cite{govindarajuo-2008-high} avoids the uncoalesced memory accesses by combining multiple thread blocks into one thread block.
Figure~\ref{fig:coalescing} shows an example of the first radix-$N1$ NTT (Kernel-1) while performing an $N$-point NTT, where $N=64$ and $N1=4$. 
In Figure~\ref{fig:coalescing}(a), 75\% of the data in the memory transaction is wasted; each thread uses only 8 bytes out of 32 bytes.
However, in Figure~\ref{fig:coalescing}(b), by combining the four-thread blocks into one and rearranging the threads such that each thread performs adjacent NTT, the data in each memory transaction is not wasted.

\begin{figure}[!tb]
  \center
  \subfloat{\includegraphics[height=1.4in]{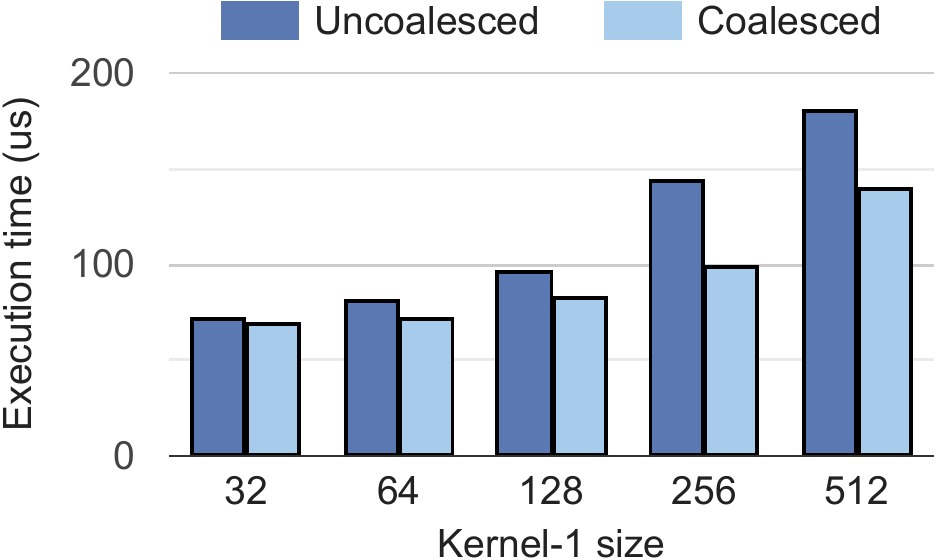}}
  \vspace{0.0in}
  \caption{
   Execution time of Kernel-1 of the SMEM implementation of NTT with and without coalesced global memory accesses in different radices when $N=2^{17}$.
   }
  \label{fig:align}
  \vspace{-0.05in}
\end{figure}

Figure~\ref{fig:align} shows the performance of Kernel-1 for the SMEM implementation of NTT with and without uncoalesced memory accesses in different radices when $N=2^{17}$.
Here, the SMEM implementation performs R (the radix of Kernel-1) point NTT by means of single block-level synchronization between R1-point NTT and R2-point NTT identically to how the process operates in Figure~\ref{fig:smem_impl}.
A larger value between R1 and R2 determines the register usage per thread, which affects the kernel occupancy.
To minimize register usage per thread, we set R1 as the smallest power of two among the numbers greater than or equal to $\sqrt{R}$ and set R2 $R/R1$.
The allocated shared memory per thread block is R1 $\times$ (the number of threads in a thread block) $\times$ 8 bytes in size.
SMEM can store and redistribute all input data processed by a block at once.
When loading input data from global memory, an uncoalesced case is established such that it has no consecutive data transactions within any single warp, with the coalescing case being the opposite of this.
By removing uncoalesced memory accesses through combining multiple thread blocks, we speed up the process by 21.6\% on average.

\begin{figure}[!tb]
  \center
	  \subfloat{\includegraphics[width=3.2in]{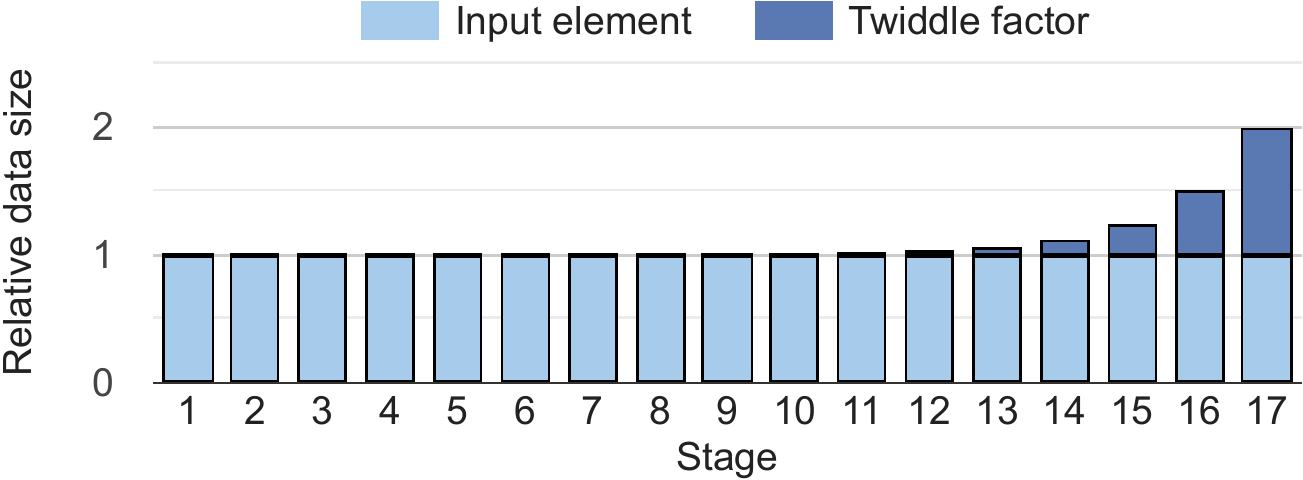}}
  \vspace{0.0in}
  \caption{The relative size of the precomputed table and input data for each stage in the radix-2 NTT implementation.}
  \label{fig:stage_data}
  \vspace{-0.05in}
\end{figure}

\textbf{Storing the precomputed table in SMEM at the early stages:}
Figure~\ref{fig:stage_data} shows the relative size of the input data and the precomputed table required at each radix-2 NTT stage. 
During the early stages, we can store a small precomputed table in SMEM, as the precomputed tables fit into SMEM. 
This can improve the cache behavior of GPUs in the early stages. 

\begin{figure}[!tb]
  \center
  \subfloat{\includegraphics[height=1.4in]{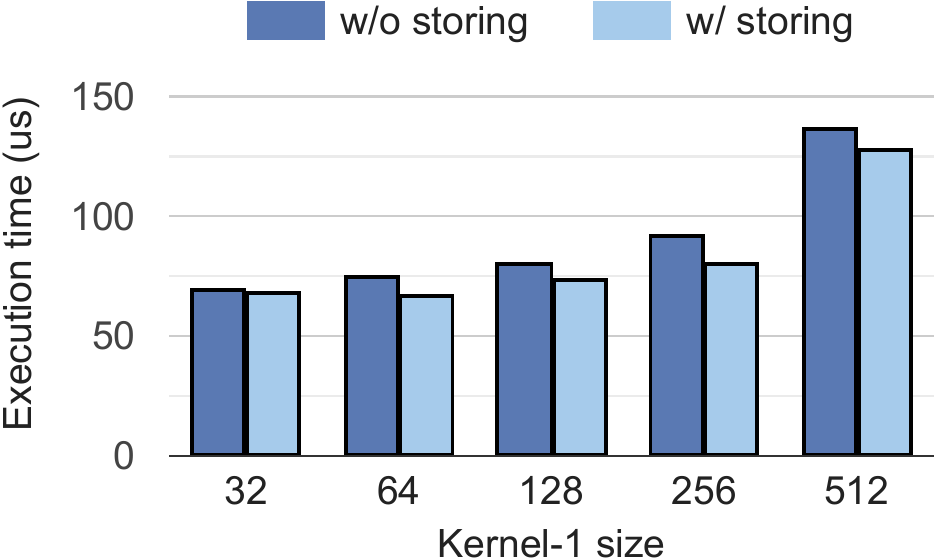}}
  \vspace{0.0in}
  \caption{
    Performance of Kernel-1 with and without storing the precomputed table in SMEM. We use the parameters ($N$, $np$) as ($2^{17}$, 21).
   }
  \label{fig:pin}
  \vspace{-0.05in}
\end{figure}

Figure~\ref{fig:pin} shows the performance with and without the precomputed table stored in SMEM on Kernel-1 in the SMEM implementation. 
Here, the base configuration is identical to when coalescing is utilized.
The storage case preloads the required twiddle factors into SMEM before performing the first R1-point NTT and then performs the NTT operation using the preloaded data.
The shared memory space allocated to the block for twiddle factors is R$\times$8 bytes in size, allowing it to load all of the twiddle factors necessary to process R-point NTT from GMEM at once during the preload phase.
The w/o storing case loads the twiddle factors directly from GMEM.
We gain a speedup of 8.4\% on average.

\begin{figure}[!tb]
  \center
  \subfloat{\includegraphics[height=3.3in]{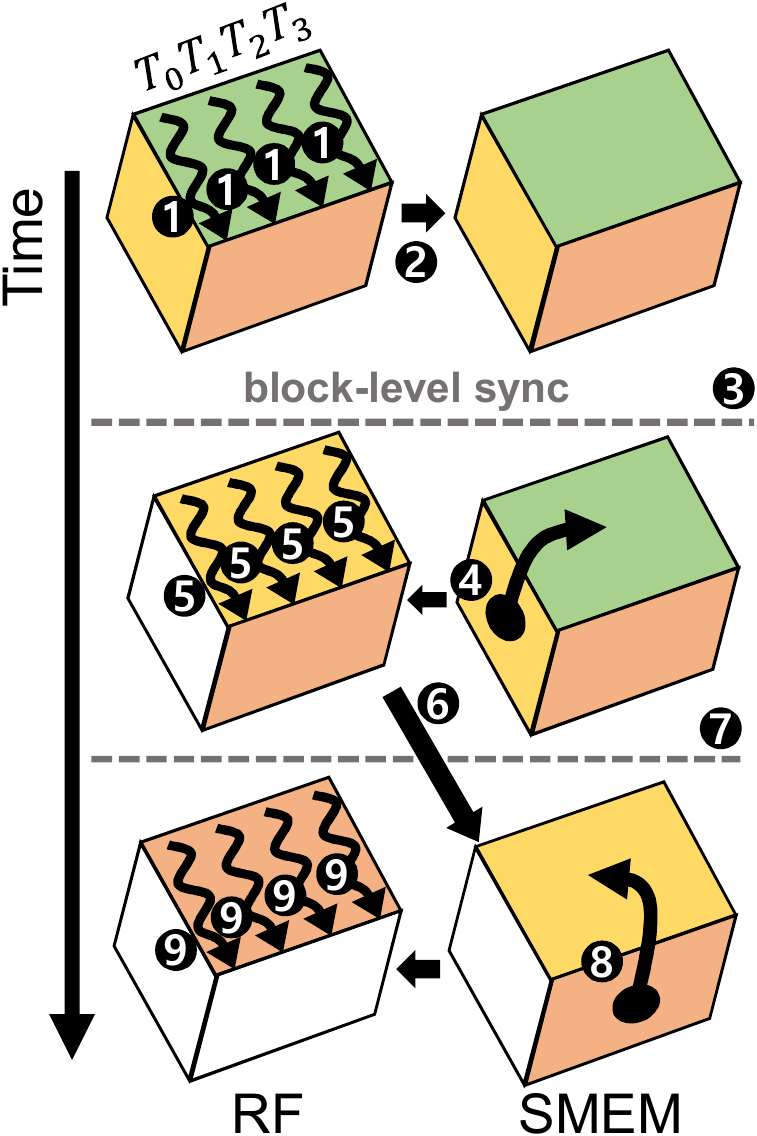}}
  \vspace{0.0in}
  \caption{
    Example of the SMEM implementation of radix-64 (4$\times$4$\times$4) NTT. There are 16 threads in a thread block, but only 4 are shown for convenience. Each thread performs 4 per-thread NTT (\encircled{1}, \encircled{5}, \encircled{9}) with data stored in the register file (RF). The outputs are stored into SMEM ({\encircled{2}}, {\encircled{6}}), followed by a block-level synchronization ({\encircled{3}}, {\encircled{7}}). Then, each thread loads the data in a transposed fashion ({\encircled{4}}, {\encircled{8}}) to run 4 per-thread NTT ({\encircled{5}}, {\encircled{9}}). Compared to a radix-8$\times$8 case in Figure~\ref{fig:smem_impl}, one more block-level synchronization is added.  
   }
  \label{fig:3d_smem}
  \vspace{-0.08in}
\end{figure}

\textbf{Trade-off between the number of block-level synchronizations and register usage:}
For the SMEM implementation, we can perform the same radix while varying the per-thread NTT sizes.
For example, to perform a radix-64 NTT, a single synchronization step is needed with 8-point-per-thread NTT when using the method depicted in Figure~\ref{fig:smem_impl}.
However, we can also handle the same radix size with 4-point-per-thread NTT and two synchronizations, as depicted in Figure~\ref{fig:3d_smem},
which is equivalent to the difference between 2D-FFT and 3D-FFT.

There is a trade-off between register usage and the number of block-level synchronizations when varying the size of the per-thread NTT.
Utilizing small-point per-thread NTT reduces the register usage for input data from $O(\sqrt{R})$ to $O(\sqrt[3]{R})$ with identically sized radix $R$. 
However, such a reduction increases the required number of synchronizations.
For example, to perform a radix-512 NTT, two synchronization steps are needed with 8-point-per-thread NTT whereas eight are needed with 2-point-per-thread NTT. 

\begin{figure}[!tb]
  \center
  \subfloat[Exeuction time of NTT]{\includegraphics[width=2.45in]{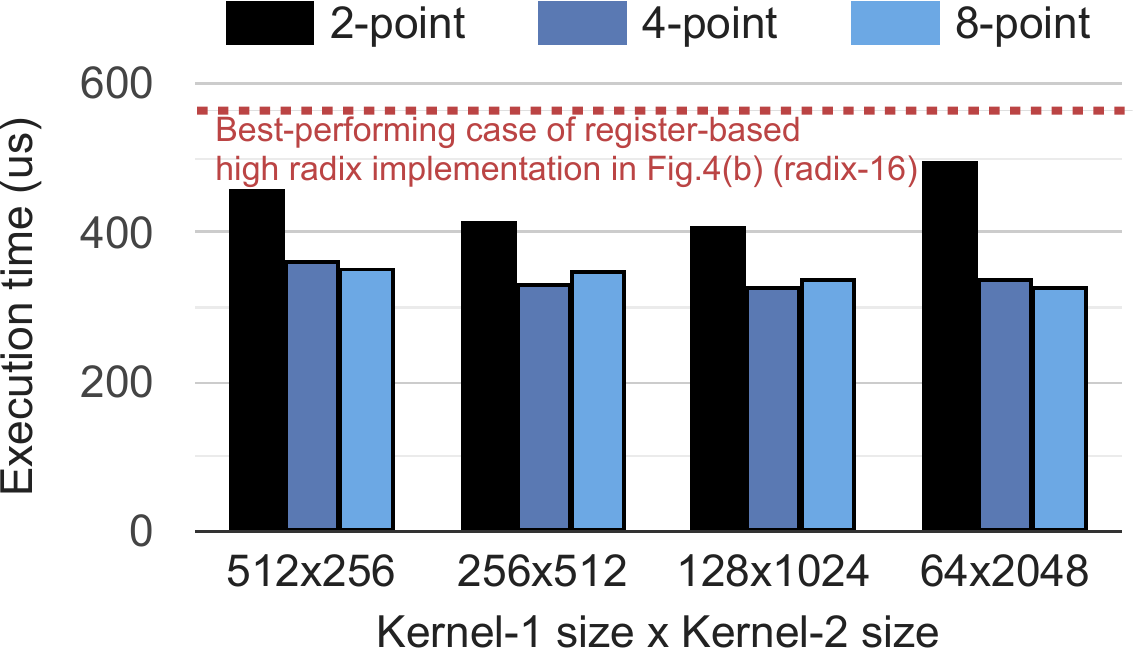}}
  \hspace{0.1in}
  \subfloat[Execution time of DFT]{\includegraphics[width=2.45in]{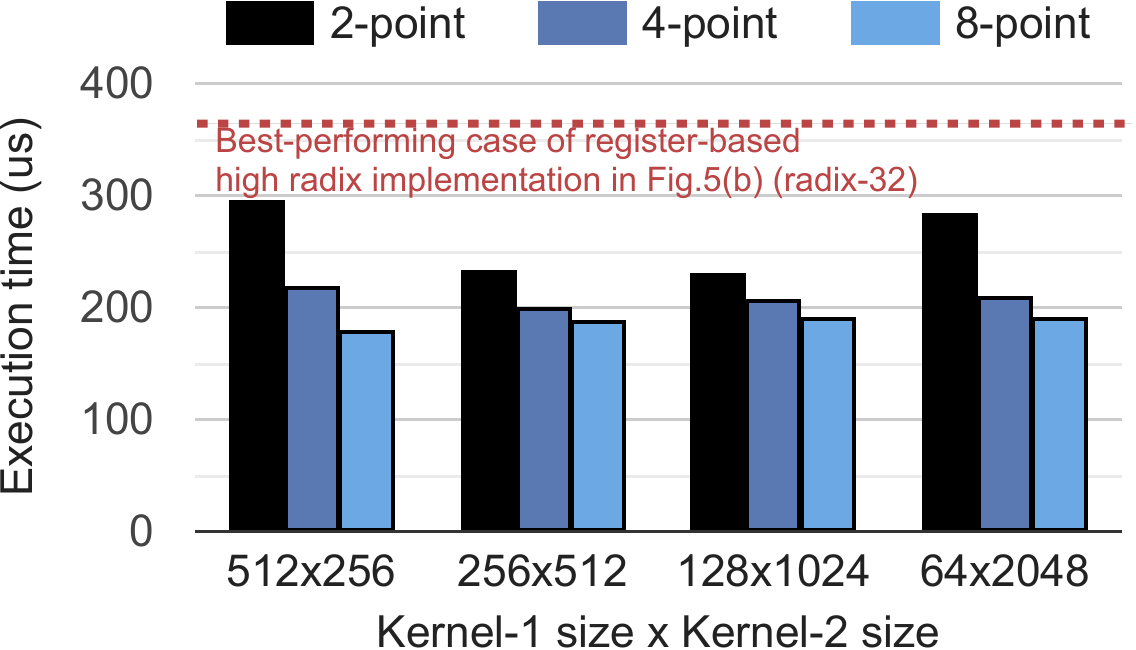}}
  \hspace{0.1in}
  \subfloat[Exeuction time of NTT w/ OT]{\includegraphics[width=2.28in]{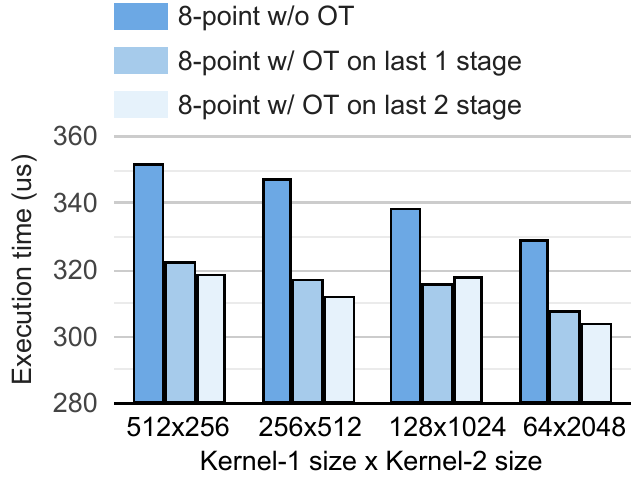}}
  \caption{
     The execution time of (a) NTT and (b) DFT in different sizes of per-thread NTT/DFT 
     and (c) NTT when on-the-fly twiddling (OT) is not applied or applied to either the last only or the last two stages 
     when 
     ($N$, $np$) = ($2^{17}$, 21). }
  \label{fig:3point}
  \vspace{-0.05in}
\end{figure}

Figure~\ref{fig:3point}(a) shows the performance outcomes of Kernel-1 and Kernel-2 with the different of per-thread NTT sizes.
R1 in the `R1-point' label refers to the size of the per-thread NTT described above. 
The execution process of the individual kernel depends on whether the size of a kernel R is or is not precisely the powers of R1.
If so, a kernel of size R performs $log_{R1}(R)$ instances of R1-point per-thread NTT and block-level synchronization between each per-thread NTT. 
If not, for the largest integer k less than $log_{R1}(R)$, a kernel of size R performs R1-point NTT by k times and $R/R1^{k}$-point NTT once at the last step.
Both Kernel-1 and Kernel-2 also perform block-level synchronization between each per-thread NTT as before.
The size of the SMEM space used for the input data is R1 $\times$ (number of threads per thread block) $\times$ 8 bytes. 
Preloading of the twiddle factors is only used in Kernel-1.
The per-thread NTT of size 4 performs 30.1\% better than that of size 2.
The per-thread NTTs of size 4 and 8 perform similarly.
The red dotted line shows the performance of the best-performing case of the register-based high radix implementation (radix-16, which takes 566 us in Figure~\ref{fig:radix_ntt}(b)).
All configurations using SMEM outperform performance than the register-based high-radix implementation.

Figure~\ref{fig:3point}(b) shows the performance of Kernel-1 and Kernel-2 in DFT with different per-thread DFT sizes. 
The configuration of each label is identical to that used for labels with the same name in NTT.
The per-thread DFT of size 2 is the slowest, as in NTT (Figure~\ref{fig:3point}(a)), and that with a size of 8 performs better compared to when a size of 4 is used.
All of the configurations of SMEM implementation on DFT also outperform the best-performing case of register-based implementation (radix-32, which takes 364.2 us, as shown in Figure~\ref{fig:radix_dft}(b)).

\section{Accelerating NTT using On-the-fly Twiddling (OT)}
\label{sec:contribution-4}
As NTT requires a modulo operation while calculating each twiddle factor, it is expensive to generate the twiddle factors in an on-the-fly manner. 
Moreover, a precomputed variable $\bar{w}$ in Shoup's modmul (Algo.~\ref{algo:shoup}) should also be calculated for each twiddle factor. 
Therefore, previous works~\cite{govindarajuo-2008-high} that generate the twiddle factors on the fly are not suitable for NTT.

We propose a novel on-the-fly twiddling (OT) technique that avoids modulo operations during the on-the-fly generation of twiddle factors. 
OT reduces the precomputed table size and thus requires fewer main memory accesses, 
alleviating the memory bandwidth bottleneck.

OT does not calculate a twiddle factor; in fact, it multiplies an input with the existing twiddle factors in the precomputed table in a consecutive fashion using the associative law: 
for the given twiddle factors ($w1$, $w2$) and an input ($x$), instead of calculating $w = w2 \times w1$ and multiplying it by $x$ as $w \times x$,
OT first calculates $x’ = w1 \times x$ first, followed by $w \times x = w2 \times x'$.

Similar to native modulo operations, OT adds a 64-bit modular multiplication for each generation of a twiddle factor. 
However, OT avoids the cost of a na\"ive modulo operation and does not calculate a new $\bar{w}$ corresponding to $w$; 
only $\bar{w1}$ and $\bar{w2}$ are needed when multiplying $w1$ and $w2$ by the input consequently.

Because the twiddle factor $w$ can be factorized recursively, there exists a tradeoff between the precomputed table size and the number of modular multiplications. 
For example, if divided into the base-2, an $N$-point NTT requires $log_2{N}$ twiddle factors, while the required number of modular multiplications for the generation of the twiddle factor is as high as $log_2{N}$ 
(e.g., for an 8-point NTT, three twiddle factors are required: $w^7 \times x = w^4 \times w^2 \times w^1 \times x$). 
From our experiments, we found that dividing into base-1024 performs best.
If the parameter $N$ is $2^{17}$, the number of the precomputed twiddle factors becomes $1024 + \frac{2^{17}}{1024}$ with base-1024.

As the number of twiddle factors required is small at the early stages for NTT, it is feasible to apply OT only to the later stages.
Figure~\ref{fig:3point}(c) shows the relationship between the number of stages that apply OT and the performance when OT is applied on the 8-point-per-thread NTT.
If OT is applied to the last two stages, the performance improves in general, but it deteriorates when the sizes of Kernel-1 and Kernel-2 are 128 and 1024, respectively, compared to cases in which OT is applied to the last stage only.

\section{Comparing the Effectiveness of the Overall Optimizations for Various Parameter Sets}

\label{sec:evaluation}
\begin{figure}[!tb]
  \center
  \subfloat{\includegraphics[height=1.4in]{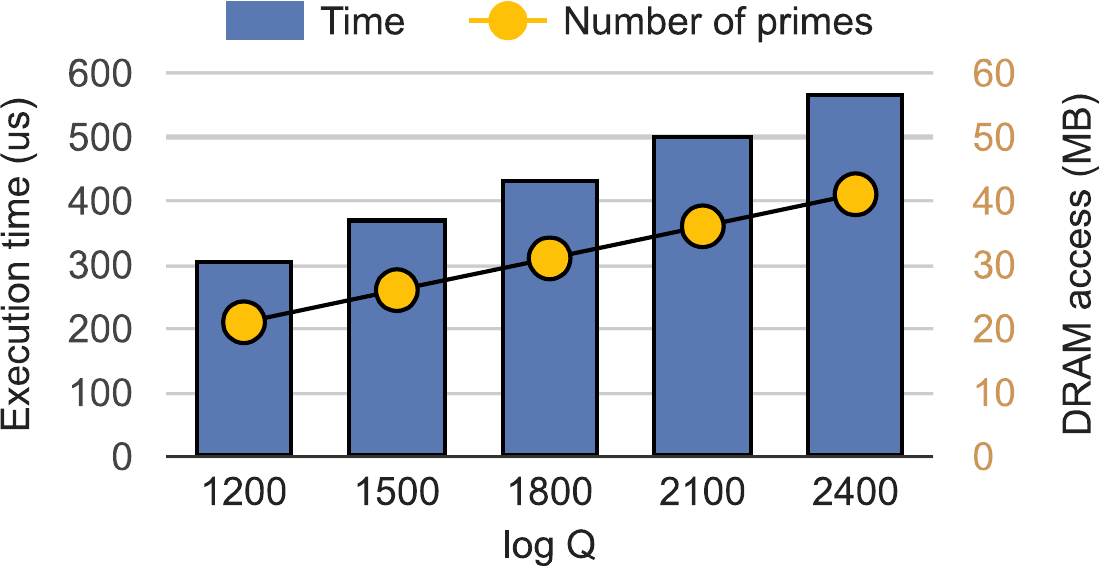}}
  \vspace{0.0in}
  \caption{
    The execution time of the SMEM implementation for NTT with the best-performing combination of radices when $N$ is $2^{17}$ in various batch sizes ($np$). The values of logQ, each corresponding to $np$, are presented.
  }
  \label{fig:logq}
  \vspace{-0.00in}
\end{figure}

We analyzed the performance impact of FFT-based optimizations in the previous sections over various bootstrappable HE parameters.
First, the performance of NTT for a single prime becomes saturated when the batch sizes exceed moderate levels. 
Figure~\ref{fig:logq} shows the performance of NTT when batching is applied with the best-performing combination of radices of Kernel-1 and Kernel-2 in the SMEM implementation across a range of ciphertext moduli (Q). 
Because the batch size of 21 already highly utilizes the GPU (see Section~\ref{sec:contribution-2}), the execution time increases linearly with the batch size.
\begin{figure}[t]
  \center
  \subfloat[Execution time of NTT w/o OT]{\includegraphics[width=3.0in]{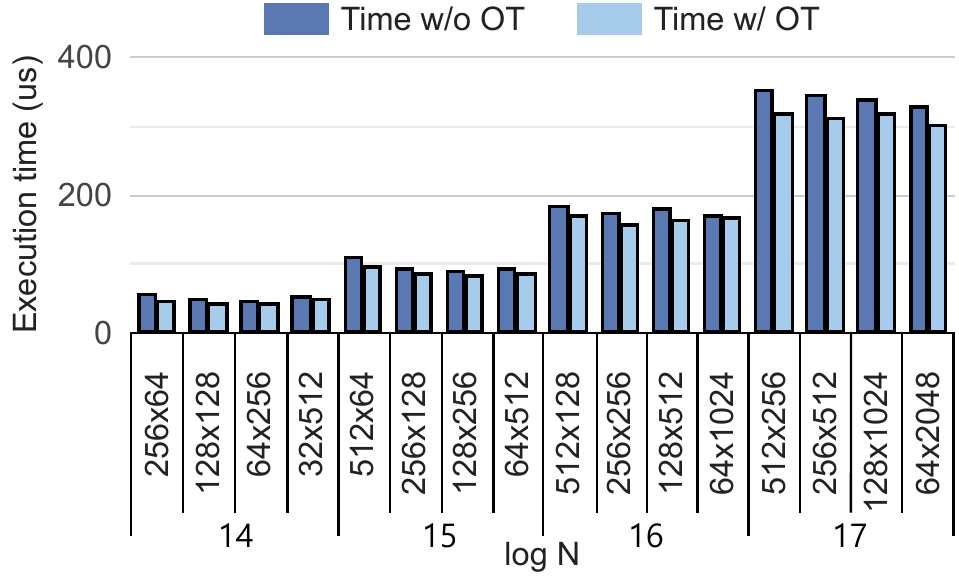}}
  \hspace{0.1in}
  \subfloat[DRAM utilization and speedup]{\includegraphics[width=2.7in]{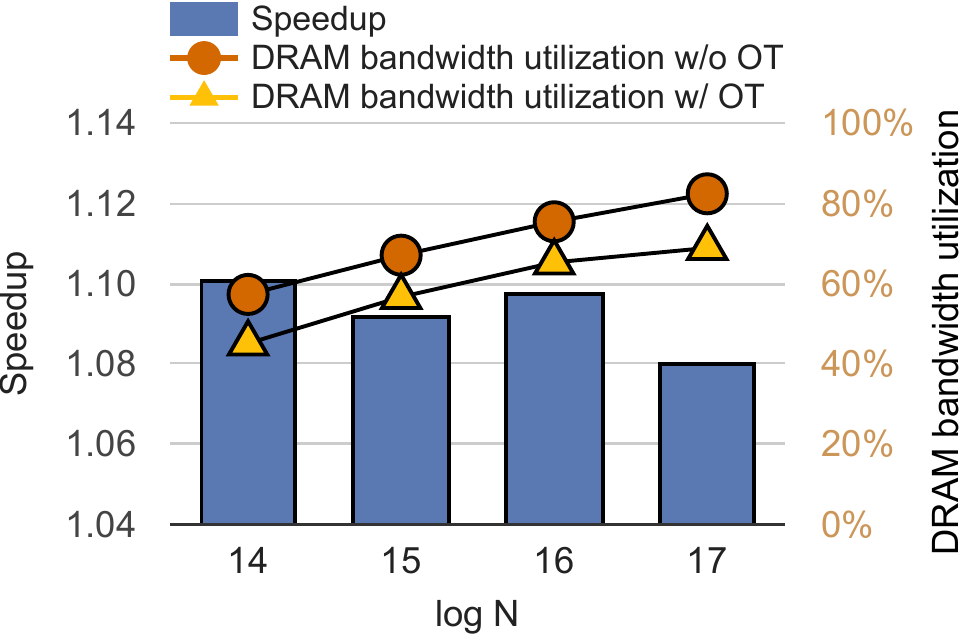}}
  \hspace{0.1in}
  \subfloat[DRAM access count]{\includegraphics[width=2.4in]{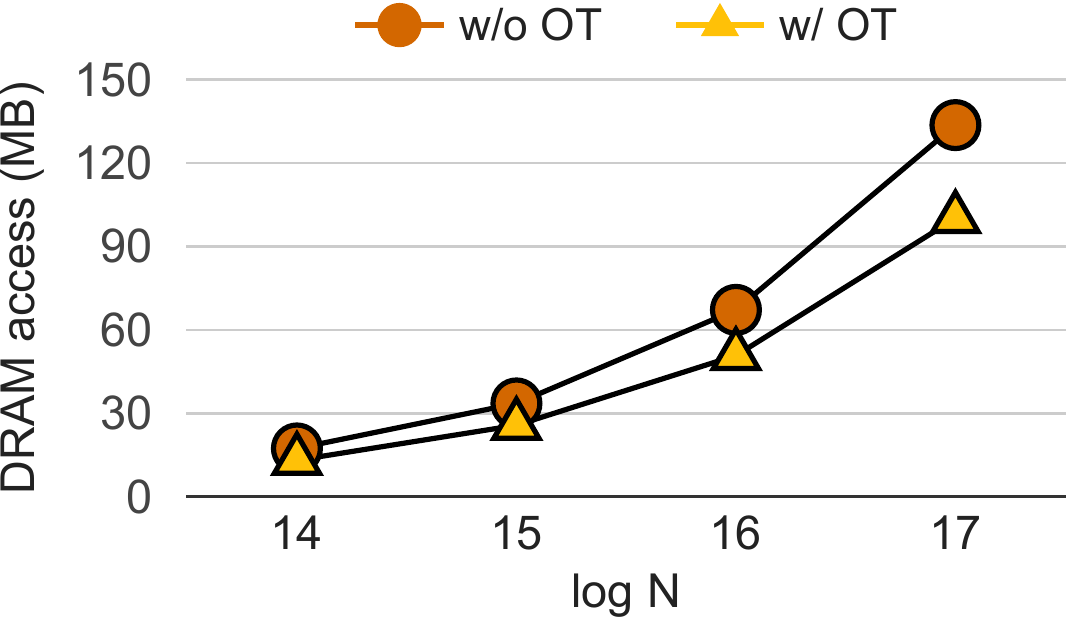}}
  \caption{
    (a) The performance of the SMEM implementation in different combinations of radices of Kernel-1 and Kernel-2 with OT and 
    (b) DRAM bandwidth utilization and performance and (c) the amount of DRAM memory access of the best-performing SMEM implementation of NTT with and without OT when $np$ is 21.
  }
  \label{fig:ntt_with_various}
  \vspace{-0.05in}
\end{figure}

Second, the performance difference between the combinations of the radices of Kernel-1 and Kernel-2 in the SMEM implementation for a given $N$ is negligible. 
Figure~\ref{fig:ntt_with_various}(a) shows the performance results for Kernel-1 and Kernel-2 across various combinations of radices and $N$. 
Similar to Figure~\ref{fig:3point}(c), the baseline configuration of the SMEM implementation is set to 8-point-per-thread NTT.
When $log{N}$ is 16, 15, and 14, the performance differences are up to 7.5\%, 15.7\%, and 16.3\%, respectively.
The trend is similar even after applying OT.

Third, OT alleviates the main-memory bandwidth bottleneck in NTT by reducing the number of GMEM accesses. 
Figure~\ref{fig:ntt_with_various}(b) and Figure~\ref{fig:ntt_with_various}(c) represent the number of DRAM accesses, the DRAM bandwidth utilization, and the performance result of the best-performing case of the SMEM implementation of NTT with and without OT. 
OT reduces the number of DRAM accesses by 24.5\%, 23.5\%, 24.5\%, and 25.1\% when N is $2^{14}$, $2^{15}$, $2^{16}$, and $2^{17}$, respectively, resulting in a DRAM bandwidth utilization reduction of 16.7\% and a speedup of 9.3\% on average.
Table~\ref{tab:logN} summarizes the NTT performance of radix-2 and the SMEM implementation with the best-performing combination of radices with and without OT.
OT results in speedups of 8.1\%, 9.8\%, 9.2\%, and 10.1\% compared to the best-performing configurations without OT when $log{N}$ is 17, 16, 15, and 14, respectively.
Compared to the radix-2 implementation results, the degree of performance improvement in the SMEM implementation with OT is by 4.2$\times$ on average.

\begin{table}[tb!]
  \centering
  \caption{The execution time of NTT with radix-2 and the SMEM implementation with and without OT for various $N$ values.}  
  \small
  \begin{tabular}{lcccccc}
  \toprule
  &&\multicolumn{5}{c}{Execution time (us) \& Speedup}\\
  $log{N}$& $np$ & \textbf{Radix}-2 &\multicolumn{2}{c}{\textbf{SMEM} w/o OT}& \multicolumn{2}{c}{\textbf{SMEM} w/ OT} \\
  \midrule
  $14$ &  21  & 166  & 48.6&[3.4$\times$] & 44.1&[3.8$\times$]  \\
  $15$ &  21  & 340  & 92.0&[3.7$\times$] & 84.2&[4.0$\times$]  \\
  $16$ &  21  & 693  & 171.8&[4.0$\times$] & 156.3&[4.4$\times$]  \\
  $17$ &  21  & 1427 & 329.0&[4.3$\times$] & 304.2&[4.7$\times$]  \\
  \bottomrule
  \end{tabular}
  \label{tab:logN}
\end{table}

We also compared our results with a prior work~\cite{kim-2020-hardware}, which accelerates NTT for large bootstrappable HE parameter sets. 
The parameter sets used in this comparison are identical.
Our OT with the SMEM implementation outperforms \cite{kim-2020-hardware} by 6.56$\times$ and 6.48$\times$ for ($N=2^{17},np=36$) and ($N=2^{17},np=42$), respectively.

\section{Related Work}
\label{sec:relatedwork}
\noindent
\textbf{Studies exploiting FPGAs for HE:}
Several prior works have attempted to accelerate NTT on FPGAs~\cite{riazi-2020-HEAX, kim-2020-hardware}. 
However, as opposed to our work, \cite{riazi-2020-HEAX} did not report the large parameter sizes required for bootstrapping operations, 
which is critical when running sophisticated, real-world applications. 
\cite{kim-2020-hardware} attempted to generate several of the twiddle factors on-the-fly with a pipeline-oriented implementation. 
However, this approach is not suitable for GPUs that exploit massive thread-level parallelism.

\noindent
\textbf{Studies exploiting GPUs for HE:} 
NTT has ample parallelism that can be exploited by a popular parallel hardware platform, GPU. 
A number of early attempts~\cite{iccis-2015-cuhe, iacr-2018-fvgpu} tried to accelerate NTT on GPUs. 
In \cite{iccis-2015-cuhe}, a special prime was exploited, called the \emph{Solinas prime}, to simplify modular multiplications.
Different moduli were used for CRT to generate multiple polynomials of residual numbers, but the methods required the sharing of a single Solinas prime while performing NTT
on these polynomials.
However, using one Solinas prime $p$ requires each modulus of CRT to be less than $\sqrt{p/(2N)}$,
significantly restricting the parameter choices.
Moreover, the design space was not explored, and the method lacked a rigorous analysis of performance limiting factors on various NTT implementations.

\section{Conclusion}
\label{sec:conclusion}
In this paper, we conducted an in-depth analysis of the algorithmic 
differences between NTT and DFT.
We applied optimizations that were
originally targeted for DFT, in this case batching, register-based high-radix 
implementation, and shared memory implementation, to NTT, a primary
component of Homomorphic Encryption (HE).
We explored the design space of NTT and determined the primary 
performance limiting factors.
Through a comprehensive analysis and design space exploration, 
we found that the main-memory bandwidth still causes a  bottleneck in
the NTT case even
after applying the aforementioned DFT optimizations.
We then proposed a novel on-the-fly technique by which to generate 
twiddle factors to alleviate the main-memory bandwidth bottleneck, leading
to an additional speedup of 9.3\% on average over typical
NTT parameters.

\section*{Acknowledgments}
This work was supported by Institute of Information \& communications Technology 
Planning \& Evaluation (IITP) grant funded by the Korea government (MSIT) 
(No. 2020-0-00840, Development and Library Implementation of Fully Homomorphic Machine
Learning Algorithms supporting Neural Network Learning over Encrypted Data) and
by the National Research Foundation of Korea (NRF) grant funded by MSIT (No. 2020R1A2C2010601).
The authors thank Dr. Eojin Lee for his valuable feedback.
Jung Ho Ahn is the corresponding author.


\balance
\bibliographystyle{IEEEtranS}
\bibliography{ref}

\begin{thebibliography}{10}
\providecommand{\url}[1]{#1}
\csname url@samestyle\endcsname
\providecommand{\newblock}{\relax}
\providecommand{\bibinfo}[2]{#2}
\providecommand{\BIBentrySTDinterwordspacing}{\spaceskip=0pt\relax}
\providecommand{\BIBentryALTinterwordstretchfactor}{4}
\providecommand{\BIBentryALTinterwordspacing}{\spaceskip=\fontdimen2\font plus
\BIBentryALTinterwordstretchfactor\fontdimen3\font minus
  \fontdimen4\font\relax}
\providecommand{\BIBforeignlanguage}[2]{{%
\expandafter\ifx\csname l@#1\endcsname\relax
\typeout{** WARNING: IEEEtranS.bst: No hyphenation pattern has been}%
\typeout{** loaded for the language `#1'. Using the pattern for}%
\typeout{** the default language instead.}%
\else
\language=\csname l@#1\endcsname
\fi
#2}}
\providecommand{\BIBdecl}{\relax}
\BIBdecl

\bibitem{aguilar-2016-nfllib}
C.~Aguilar-Melchor, J.~Barrier, S.~Guelton, A.~Guinet, M.~Killijian, and
  T.~Lepoint, ``{NFLlib: NTT-based Fast Lattice Library},'' in
  \emph{Cryptographers’ Track at the RSA Conference}, 2016.

\bibitem{iacr-2018-fvgpu}
A.~A. Badawi, B.~Veeravalli, C.~F. Mun, and K.~M.~M. Aung, ``{High-Performance
  FV Somewhat Homomorphic Encryption on GPUs: An Implementation Using CUDA},''
  \emph{The International Association for Cryptologic Research Transactions on
  Cryptographic Hardware and Embedded Systems}, vol. 2018, no.~2, 2018.

\bibitem{bajard-2016-full}
J.~Bajard, J.~Eynard, M.~A. Hasan, and V.~Zucca, ``{A Full RNS Variant of FV
  like Somewhat Homomorphic Encryption Schemes},'' in \emph{International
  Conference on Selected Areas in Cryptography}, 2016.

\bibitem{barrett-1986-implementing}
P.~Barrett, ``{Implementing the Rivest Shamir and Adleman Public Key Encryption
  Algorithm on a Standard Digital Signal Processor},'' in \emph{Conference on
  the Theory and Application of Cryptographic Techniques}, 1986.

\bibitem{bell-2009-implementing}
N.~Bell and M.~Garland, ``{Implementing Sparse Matrix-Vector Multiplication on
  Throughput-Oriented Processors},'' in \emph{Proceedings of the Conference on
  High Performance Computing Networking, Storage and Analysis}, 2009.

\bibitem{chang-2016-accelerating}
B.~Chang, B.~Goi, R.~C.~W. Phan, and W.~Lee, ``{Accelerating Multiple Precision
  Multiplication in GPU With Kepler Architecture},'' in \emph{2016 IEEE 18th
  International Conference on High Performance Computing and Communications},
  2016.

\bibitem{icfcds-2017-seal}
H.~Chen, K.~Laine, and R.~Player, ``{Simple Encrypted Arithmetic Library-SEAL
  v2.1},'' in \emph{International Conference on Financial Cryptography and Data
  Security}, 2017.

\bibitem{cheon-2018-full}
J.~H. Cheon, K.~Han, A.~Kim, M.~Kim, and Y.~Song, ``{A Full RNS Variant of
  Approximate Homomorphic Encryption},'' in \emph{International Conference on
  Selected Areas in Cryptography}, 2018.

\bibitem{cheon2018bootstrapping}
J.~H. Cheon, K.~Han, A.~Kim, M.~Kim, and Y.~Song, ``{Bootstrapping for
  Approximate Homomorphic Encryption},'' in \emph{Annual International
  Conference on the Theory and Applications of Cryptographic Techniques}.\hskip
  1em plus 0.5em minus 0.4em\relax Springer, 2018, pp. 360--384.

\bibitem{cochran-1967-fast}
W.~T. Cochran, J.~W. Cooley, D.~L. Favin, H.~D. Helms, R.~A. Kaenel, W.~W.
  Lang, G.~C. Maling, D.~E. Nelson, C.~M. Rader, and P.~D. Welch, ``{What is
  the Fast Fourier Transform?}'' \emph{Proceedings of the IEEE}, vol.~55,
  no.~10, 1967.

\bibitem{mc-1965-ntt}
J.~W. Cooley and J.~W. Tukey, ``{An Algorithm for the Machine Calculation of
  Complex Fourier Series},'' \emph{Mathematics of Computation}, vol.~19,
  no.~90, 1965.

\bibitem{iccis-2015-cuhe}
W.~Dai and B.~Sunar, ``{cuHE: A Homomorphic Encryption Accelerator Library},''
  in \emph{International Conference on Cryptography and Information Security},
  2015.

\bibitem{goey-2020-accelerating}
J.~Goey, W.~Lee, B.~Goi, and W.~Yap, ``{Accelerating Number Theoretic Transform
  in GPU Platform for Fully Homomorphic Encryption},'' \emph{The Journal of
  Supercomputing}, 2020.

\bibitem{govindarajuo-2008-high}
N.~K. Govindaraju, B.~Lloyd, Y.~Dotsenko, B.~Smith, and J.~Manferdelli, ``{High
  Performance Discrete Fourier Transforms on Graphics Processors},'' in
  \emph{Proceedings of the 2008 ACM/IEEE Conference on Supercomputing}, 2008.

\bibitem{allen-2009-computational}
E.~Gutierrez, S.~Romero, M.~A. Trenas, and O.~Plata, ``{Experiences with
  Mapping Non-linear Memory Access Patterns into GPUs},'' in
  \emph{International Conference on Computational Science}, 2009.

\bibitem{han-2020-better}
K.~Han and D.~Ki, ``{Better Bootstrapping for Approximate Homomorphic
  Encryption},'' in \emph{Cryptographers’ Track at the RSA Conference}, 2020.

\bibitem{jsc-2014-harvey}
D.~Harvey, ``{Faster Arithmetic for Number-Theoretic Transforms},''
  \emph{Journal of Symbolic Computation}, vol.~60, 2014.

\bibitem{jia-2018-dissecting}
Z.~Jia, M.~Maggioni, B.~Staiger, and D.~P. Scarpazza, ``{Dissecting the NVIDIA
  Volta GPU Architecture via Microbenchmarking},'' \emph{Computing Research
  Repository}, 2018.

\bibitem{jung-2020-heaan}
W.~Jung, E.~Lee, S.~Kim, K.~Lee, N.~Kim, C.~Min, J.~H. Cheon, and J.~Ahn,
  ``{HEAAN Demystified: Accelerating Fully Homomorphic Encryption Through
  Architecture-centric Analysis and Optimization},'' \emph{arXiv preprint
  arXiv:2003.04510}, 2020.

\bibitem{kim-2020-hardware}
S.~Kim, K.~Lee, W.~Cho, Y.~Nam, J.~H. Cheon, and R.~A. Rutenbar, ``{Hardware
  Architecture of a Number Theoretic Transform for a Bootstrappable RNS-based
  Homomorphic Encryption Scheme},'' in \emph{2020 IEEE 28th Annual
  International Symposium on Field-Programmable Custom Computing Machines
  (FCCM)}, 2020.

\bibitem{kopcke-2019-generating}
B.~K{\"o}pcke, M.~Steuwer, and S.~Gorlatch, ``{Generating Efficient FFT GPU
  Code with Lift},'' in \emph{Proceedings of the 8th ACM SIGPLAN International
  Workshop on Functional High-Performance and Numerical Computing}, 2019.

\bibitem{lloyd-2008-fast}
D.~B. Lloyd, C.~Boyd, and N.~Govindaraju, ``{Fast Computation of General
  Fourier Transforms on GPUs},'' in \emph{2008 IEEE International Conference on
  Multimedia and Expo}, 2008.

\bibitem{nukada-2008-bandwidth}
A.~Nukada, Y.~Ogata, T.~Endo, and S.~Matsuoka, ``{Bandwidth Intensive 3-D FFT
  Kernel for GPUs Using CUDA},'' in \emph{Proceedings of the 2008 ACM/IEEE
  conference on Supercomputing}, 2008.

\bibitem{tesla-2018-v100}
{NVIDIA Corporation}, ``{V100 GPU Architecture Whitepaper},'' 2017.

\bibitem{nvidia-2020-sass}
{NVIDIA Corporation}, ``{CUDA Binary Utilities},''
  \url{https://docs.nvidia.com/cuda/cuda-binary-utilities/index.html}, July
  2020.

\bibitem{nvidia-2019-ptx}
{NVIDIA Corporation}, ``{CUDA C++ Programming Guide},''
  \url{https://docs.nvidia.com/cuda/cuda-c-programming-guide/index.html}, July
  2020.

\bibitem{pop-2012-towards}
T.~P{\"o}ppelmann and T.~G{\"u}neysu, ``{Towards Efficient Arithmetic for
  Lattice-based Cryptography on Reconfigurable Hardware},'' in
  \emph{International Conference on Cryptology and Information Security in
  Latin America}, 2012.

\bibitem{poppelmann-2015-high}
T.~P{\"o}ppelmann, T.~Oder, and T.~G{\"u}neysu, ``{High-performance Ideal
  Lattice-based Cryptography on 8-bit ATxmega Microcontrollers},'' in
  \emph{International Conference on Cryptology and Information Security in
  Latin America}, 2015.

\bibitem{riazi-2020-HEAX}
M.~S. Riazi, K.~Laine, B.~Pelton, and W.~Dai, ``{HEAX: An Architecture for
  Computing on Encrypted Data},'' in \emph{Proceedings of the Twenty-Fifth
  International Conference on Architectural Support for Programming Languages
  and Operating Systems}, 2020.

\bibitem{fsc-1978-he}
R.~L. Rivest, L.~Adleman, and M.~L. Dertouzos, ``{On Data Banks and Privacy
  Homomorphisms},'' \emph{Foundations of Secure Computation}, vol.~4, no.~11,
  1978.

\bibitem{roy-2019-fpga}
S.~S. Roy, F.~Turan, K.~Jarvinen, F.~Vercauteren, and I.~Verbauwhede,
  ``{FPGA-based high-performance parallel architecture for homomorphic
  computing on encrypted data},'' in \emph{HPCA}, 2019.

\bibitem{roy-2014-compact}
S.~S. Roy, F.~Vercauteren, N.~Mentens, D.~D. Chen, and I.~Verbauwhede,
  ``{Compact Ring-LWE Cryptoprocessor},'' in \emph{International Workshop on
  Cryptographic Hardware and Embedded Systems}, 2014.

\bibitem{stvrelak-2018-performance}
D.~St{\v{r}}el{\'a}k and J.~Filipovi{\v{c}}, ``{Performance Analysis and
  Autotuning Setup of the cuFFT Library},'' in \emph{Proceedings of the 2nd
  Workshop on AutotuniNg and ADaptivity AppRoaches for Energy Efficient HPC
  Systems}, 2018.

\bibitem{ulu-2020-high}
M.~E. Ulu, ``{High Performance Number Theoretic Transforms in Cryptography},''
  Ph.D. dissertation, Middle East Technical University, 2020.

\bibitem{shoup-2001-ntl}
S.~Victor, ``{NTL: A Library for Doing Number Theory},'' \url{http://www.
  shoup. net/ntl/}, 2016.

\bibitem{zhang-2017-gpu}
F.~Zhang, C.~Hu, Q.~Yin, and W.~Hu, ``{A GPU Based Memory Optimized Parallel
  Method for FFT Implementation},'' \emph{Computer Research Repository}, 2017.

\end{thebibliography}

\end{document}